\documentclass[%
reprint,
 superscriptaddress,
 amsmath,amssymb,
pra,
floatfix,
]{revtex4-2}

\usepackage{mathtools}
\usepackage{float} 

\usepackage{lipsum} 
\usepackage[position=top]{subfig}
\usepackage{graphicx}% Include figure files
\usepackage{dcolumn}% Align table columns on decimal point
\usepackage{bm}% bold math
\usepackage[hidelinks,colorlinks=true,linkcolor=blue,citecolor=blue]{hyperref}

\usepackage{dsfont}
\usepackage{array}
\newcolumntype{P}[1]{>{\centering\arraybackslash}p{#1}}
\usepackage{subcaption}
\usepackage{graphicx}
\usepackage{adjustbox}

\usepackage{braket}
\begin{document}

\preprint{APS/123-QED}

% \title{Deflection due to Entanglement Sensing and Fundamental Humpy-Dumpy Effect: Thermal and Diffusion Requirements in Massive Stern-Gerlach Interferometry}

%\title{Continous Variables Formalism for Massive Stern-Gerlach Interferometry: Humpy-Dumpy Effect, Thermal and Diffusion Requirements for Entanglement Sensing}

%\title{General Gaussian Formalism for Massive Stern-Gerlach Interferometry: Humpy-Dumpy Effect, Thermal and Diffusion Requirements for Entanglement Sensing}

%\title{Mass-Independent Entanglement between Non-Gaussian Dynamical Quantum Masses}

%\title{Mass-Independent Gravitational Entanglement via Non-Gaussian Dynamics of Quantum Masses}

%\title{Mass-Independent Gravitationally Induced Entanglement}
%\title{Mass and Thermality Independent Regime of Gravitational Entanglement}
%\title{Mass-Independent Entanglement Mediated by Gravity}

%\title{Mass-Independent Scheme for Gravitationally Induced Entanglement}

\title{Mass-Independent Gravitationally Induced Entanglement}

%\title{Gaussian Entanglement between Non-Gaussian Dynamical Quantum Masses}

\author{Lorenzo Braccini} \email{lorenzo.braccini@gmail.com}
\affiliation{Department of Physics and Astronomy, University College London, Gower Street, WC1E 6BT London, United Kingdom}
%Lines break automatically or can be forced with \\

\author{Alessio Serafini}
\affiliation{Department of Physics and Astronomy, University College London, Gower Street, WC1E 6BT London, United Kingdom}

\author{Sougato Bose}
\affiliation{Department of Physics and Astronomy, University College London, Gower Street, WC1E 6BT London, United Kingdom}

%\date{\today}% It is always \today, today,
             %  but any date may be explicitly specified
%\keywords{Suggested keywords}%Use showkeys class option if keyword
                              %display desired

\begin{abstract}

We analytically solve the entangling quantum dynamics of two interacting Stern-Gerlach Interferometers~(SGI). Each SGI exploits an operator-valued force applied by a qubit to create and recombine a non-Gaussian state of matter. The entangling phase between the two qubits generated by the leading-order gravitational interaction of the massive degrees of freedom is found to be mass-independent, both for unitary and open dynamics, irrespective of the temperature and squeezing of the initial states. Further, we show that the solution of the four interferometric paths reveals that the mere presence of the interaction does not allow for a perfect recombination of the centre of mass. This second-order effect, alongside higher-order interaction terms, can be used to bound the mass from above and below, thus restricting the experiment's regime to mesoscopic masses. By solving the open dynamics which includes diffusion and dephasing with initial squeezed thermal states, the bounds are tightened by the inclusion of realistic experimental noise. We discuss diamagnetic levitated masses with embedded NV-centres as a specific physical implementation.

\end{abstract}

\maketitle

%\tableofcontents

\section{Introduction} 

%\textit{Introduction --} 
Quantum states of matter interacting via gravity could probe the non-classical nature of the gravitational field via the detection of Gravitationally Induced Entanglement (GIE)~\cite{bose_spin_2017,marletto_gravitationally-induced_2017,bose2016matter} (also dubbed Quantum Gravity Induced Entanglement of Masses \cite{van_de_kamp_quantum_2020}), an effect predicted by low energy quantum gravity~\cite{christodoulou_possibility_2019,marshman_locality_2020,carney_newton_2022,chen_quantum_2022,bose_mechanism_2022,bengyat_gravity_2023,christodoulou_locally_2022,belenchia_quantum_2018,biswas_gravitational_2022}. %Classical theories of gravity have to be quite unphysical (eg non-local) or extend the definition of classicality severely ~\cite{martin-martinez_what_2022,classical_aziz_2025,trillo_diosi_2025,angeli_positivity_2025} to allow for GIE.
%while its implications on (non-local) \textit{classical} theories of gravity remain under debate. % 
Having reached the quantum regime~\cite{ballestero_levitodynamics_2021, delic_cooling_2020, tebbenjohanns_motional_2020,magri_real_2021}, levitated Nano-Particles (NPs) represent one of the most promising systems for GIE detection. High-delocalisations of the NPs' wavefunction are required to sense GIE, as they are effective force sensors for weak gravitational effects, which could also be used to investigate collapse models and detect \mbox{particles~\cite{giovannetti_quantum_2004,Romero_Isart_quantum_2011,bassi_models_2013,geraci_sensing_2015,hempston_force_2017,tao_yectonewton_2023,fuchs_measuring_2024,bose_massive_2023,kilian_requirements_2023,wu2023quantum}.}

Gaussian protocols achieve delocalisation via squeezing of the Continuous Variable (CV) describing the centre of mass degree of freedom of the NP~\cite{jansky_strong_1992}, recently reported in experimental settings~\cite{Quantum_Kamba_2025,quantum_rossi_2025}, which may be employed to detect GIE~\cite{qvarfort2020mesoscopic,weiss_large_2021,pedernales_enhancing_2022,fujita_inverted_2023,poddubny_nonequilibrium_2025,toros_relativistic_2025}. In non-Gaussian matter-wave interferometry, linear interactions between a qubit and a generic oscillator, which could also be a NP, can be used to create and recombine a superposition of Gaussian states in the CV \cite{bose_qubit_2006}. Named Stern-Gerlach Interferometry (SGI)~\cite{margalit_realization_2018}, this behaviour is predicted for Nitrogen-Vacancy (NV) centres doping a diamagnetic NP~\cite{scala_matter_2013,yin_large_2013,wan_free_2016,marshman_constructing_2022}, superconducting flux qubits near magnetic NP~\cite{romero_isart_quantum_2012,nair_massive_2023}, and atoms coupled via optical fields to charged NPs ~\cite{toros_creating_2021}. When the NP wavefunction is recombined, the relative phase between the two interferometric paths can be detected via qubit measurements to sense forces (shown to be independent of the number of initial thermal phonons of the NP CV~\cite{scala_matter_2013}) and entanglement~\cite{bose_spin_2017,bose2025spinbasedpathwaytestingquantum}. Isolation and control of such systems have been extensively studied~\cite{pedernales_motional_2020,grossardt_2020_acceleration,tilly_qudits_2021,schut_improving_2022, braccini_large_2023,japha_quantum_2023, xiang2024phononinducedcontrastmatter, zhou2024gyroscopicstabilitynanoparticlessterngerlach}.  

However, theoretical descriptions of GIE experiments involving SGIs have been achieved only at the level of the gravitational interaction (albeit quantum
gravitational \cite{marshman_locality_2020,bose_mechanism_2022,christodoulou_locally_2022}) of the {\em superposed classical (Ehrenfest) trajectories} taken by the NPs. 
These analyses avoid a full wavefunction treatment and prevent a description of the two-body quantum dynamics~\footnote{Formally, a large initial superposition ($\Delta x$) much greater than the spread  of the wavefunction ($\delta x$) each branch allows to treat the CV state as a superposition of eigenstate of the position operators ($\hat{X}_i$), and hence eigenstates of the Newtonian potential $G M^2/|\hat{X}_1 - \hat{X}_2|$. However, this is \text{not} an eigenstate of the kinetic term $\hat{P}^2/(2M)$ and, for any initial Classical state, $\delta x>\Delta x=0$.} that incorporates state delocalisation and recombination.
\begin{figure}[!b]
\includegraphics[width=0.5\textwidth]{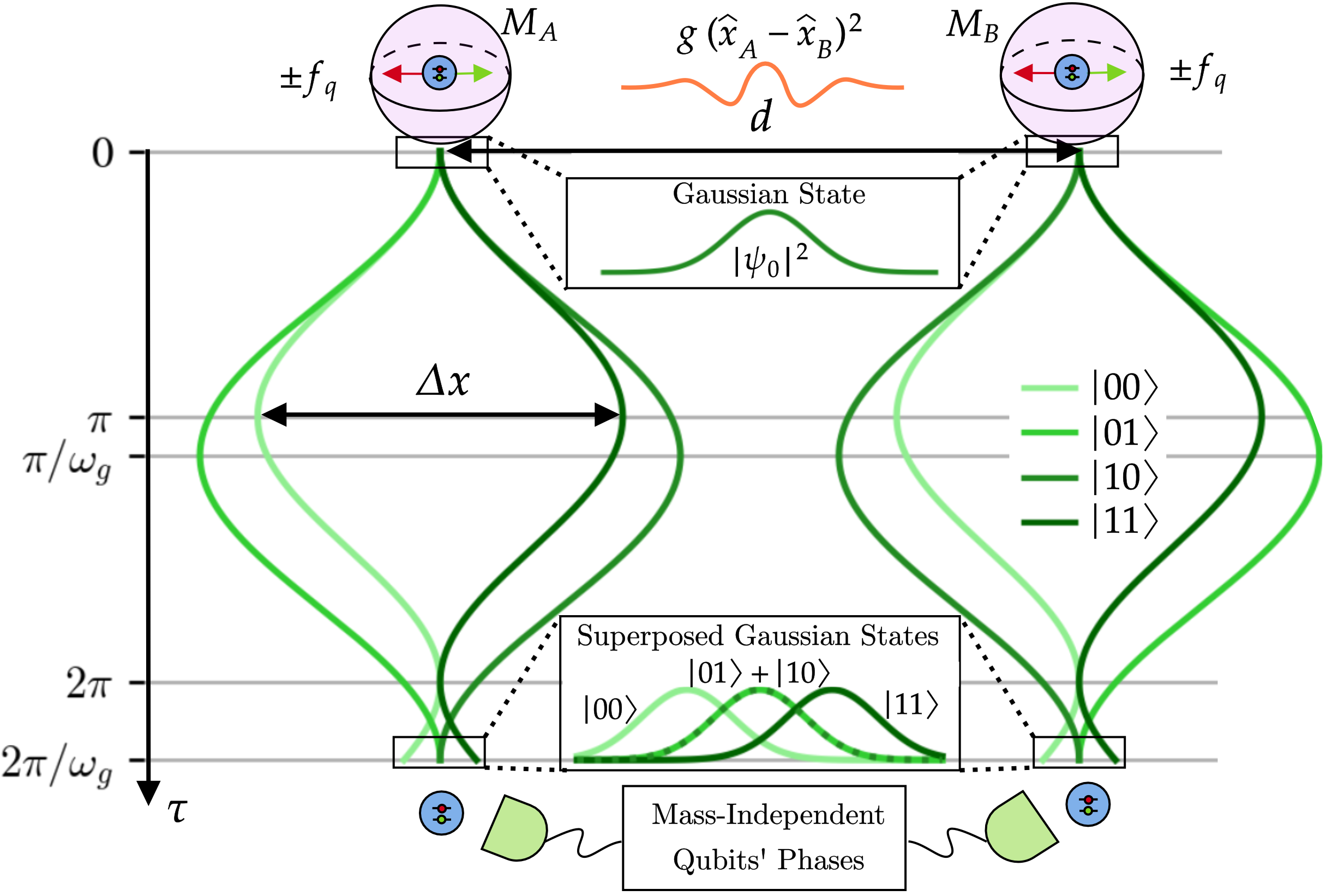}
  \caption{\label{fig:diagram} Two SGIs entangling via a Gaussian quantum interaction: the four interferometric paths of the two NPs are  in green ($\omega_g =\sqrt{1 -2g}$, $f_q = 1$, $g=0.1$).}% The common shift of the centres of the SG interferometers has been already considered.}
\end{figure}

In this work, we provide the first operational description at the level of density matrix --  necessary for any quantum experiment --  of GIE detection via non-Gaussian dynamical quantum state of matter which encompasses: (a) the process of state preparation from a classical CV Gaussian state to a highly-delocalised one, (b) the entangling open quantum dynamics during such process, and (c) the final entanglement measurements. Analytical solutions of the entangling dynamics both unitary and open can be found as a ``superposition of quantum Gaussian processes"~\cite{braccini_superposition_2025}. We rigorously show that the leading order entangling phase is \textit{independent} of the NPs' mass (in Appendix~\ref{app:consideration}, we also provide the reader with a heuristic justification of such independence based on semiclassical trajectories). The four exact interferometric paths taken by the two NPs (see Fig.~\ref{fig:diagram}) shows that a perfect recombination is not possible, but this second-order effect does not represent an experimental challenge for mesoscopic masses, while it scales with the number of phonons. The mass-independent constraint for GIE detection allows us to reduce and explore a large parameter space for the open dynamics realistic for experiment.

 \vspace{0.15cm}
\section{Mass Independence}
%\textit{Mass Independence --} 
Without considering a specific experimental implementation, in this section, we analytically treat the ideal unitary dynamics of two general SGIs entangling via a quantum Gaussian interaction, along with the detection of the leading order entanglement through qubit measurements. Two identical NPs of equal mass $M$ are in two traps of the same frequency $\omega$ centered at distance $d$ in the $y$-direction. The CV observable $\hat{X}_i$ and $\hat{P}_i$ -- with commutation relation $[ \hat{X}_j, \hat{P}_k] = i \hbar \delta_{jk} $, where $\hbar$ is the Planck constant -- describe the position and momentum in the centre of mass degree of freedom the $i$-th NP in the $x$-direction, relative to each trap centre. The degrees of freedom in the other axes will not be considered. %This choice of orientation (often named linear set up) will become clear from the following analysis.
The ground state spread of each mass is $x_0 = \sqrt{\hbar/(2 M \omega)}$. Each mass interacts with a qubit via the Stern-Gerlach interaction, for which the mass experiences an operator-valued force $F_q$ controlled by the corresponding qubit's state. The unitless Pauli operators of the $i$-th qubit are  $\hat{\sigma}_j^{(i)}$, with $j \in \{ x,y,z \}$. Thus, the Hilbert space of two SGI is $\mathcal{H} \sim  L^2(\mathds{R}^2) \otimes \mathds{C}^{4}$. In~the interaction picture with respect to the qubit energy splitting ($  \hbar \omega_q \hat{\sigma}_z/2 $), the Hamiltonian of the $i$-th SGI is~\cite{scala_matter_2013}
\begin{equation}
\label{eq:SG2}
    \hat{H}_{SG}^{(i)} =  \frac{1}{2} M \omega^2 \hat{X}_i^2 + \frac{\hat{P}_i^2}{2M}  - F_q \hat{X}_i \otimes \sigma_z^{(i)} \;.
\end{equation}
During the dynamics, given the \textit{parallel}~orientation~\cite{schut_improving_2022,braccini_large_2023}, the NPs interact through the Newtonian potential
\begin{equation}
\label{eq:Newton}
    \hat{V} =  - \frac{G M^2}{ \sqrt{(\hat{X}_1 - \hat{X}_2)^2 + d^2}} \;, %\hspace{0.5cm }\hat{V}_p =  \frac{- A}{ |(\hat{X}_1 - \hat{X}_2)^2 + d^2)|^{n/2}} 
\end{equation}
where $G$ is the Newtonian constant. The physical interaction between the two NPs is expanded to the second (leading) order in the centre of mass position of the particles, by moving to unitless operators ($\hat{x}_i = \hat{X}_i/(\sqrt{2} x_0) $ and  $\hat{p}_i = \hat{P}_i (\sqrt{2} 
x_0/\hbar) $) and assuming small displacement from the centre of the trap (see \cite{weiss_large_2021,ivan_twoqubit_2024} and Appendix \ref{app:entanglement_coupling}, where the Coulomb and Casimir interactions for different orientation are considered).  %the leading order interaction is
%\begin{equation}
%   V/(\hbar \omega) \approx - 2 g \left( \hat{x}_1 - \hat{x}_2 \right)^2, \hspace{0.2cm} \text{where} \hspace{0.2cm}  g =
%\frac{G M }{  d^3 \omega^2} \;.
%\end{equation} This expansion is performed in the  for a general $1/r^n$ interaction with an arbitrary orientation, w. 
The total Hamiltonian can be rewritten as (omitting tensor products) 
\begin{align}
\label{eq:hamiltonian_unitary}
   \frac{2 \hat{H}_g}{\hbar \omega} = \sum_{i = 1,2}  &\left[ \hat{p}_i^2 + (1-g)\hat{x}_i^2  -  f_q   \hat{x}_i \hat{\sigma}_z^{(i)}  \right]  + 2  g \hat{x}_1 \hat{x}_2,
\end{align}
where the unitless qubit-mass force and the entanglement coupling are respectively given by
\begin{equation}
f_q = \frac{F_q}{\sqrt{\hbar M \omega^3}}, \hspace{1cm} g = \frac{G M }{  d^3 \omega^2}.
\end{equation}
Thus, by moving to ground-state spread and frequency units (the time is $\tau = \omega t$), the four experimental parameters $(F_q, M, \omega, d)$ at the Gaussian order reduces to only two unit-less variables $(f_q, g)$, which uniquely define the unitary dynamics, given an initial state  ($\hat{\varrho}_0$). 

In the ideal case, each qubit is prepared in an equal uncorrelated superposition described by the density matrix $\hat{\varrho}^q_0 = \ket{+}\bra{+} \otimes \ket{+}\bra{+}$, where $\ket{+} = (\ket{1} + \ket{-1})/\sqrt{2}$ in the computational basis, and each NP is initialised in identical ground states. Two-mode CV Gaussian states find an analytical and compact description by defining the vector of the operators $\hat{r} := (\hat{x}_1, \; \hat{p}_1,\; \hat{x}_2, \; \hat{p}_2)^{\rm T}$, such that the commutation relations read $[\hat{r},\hat{r}^{\rm T}] = i \Omega = i\Omega_1 \oplus \Omega_1 $, where the symplectic form is
$\Omega_1 = \begin{pmatrix}
        0 & 1 \\
        -1 & 0 
    \end{pmatrix}$~\cite{serafini_quantum_2017}.
Displacement operators, defined as $\hat{\mathcal{D}}_{\bar{r}} := \exp( \bar{r}^{\rm T} \Omega   \hat{r})$ (where $\bar{r}  = (\bar{x}_1, \; \bar{p}_1, \; \bar{x}_2, \; \bar{p}_2)^{\rm T} \in \mathds{R}^4$ is a vector of classical variable), form an orthonormal basis for the space of CV operators providing their phase space representation. For a Gaussian quantum state~($\hat{\varrho}_{g})$, the characteristic function
used for this basis expansion ($\chi_{g} (\bar{r} ) := \text{Tr}_m [ \hat{\mathcal{D}}_{- \bar{r}}  \hat{\varrho}_{g} ] $), which is the Fourier transform of the Wigner function, is Gaussian~\cite{serafini_quantum_2017}. Hence, the initial state of the system is
\begin{equation}
\label{eq:initial_state}
    \hat{\varrho}_0 = \frac{1}{(2 \pi)^4}  \int_{\mathds{R}^{4}} \text{d} \bar{r} \; {\rm e}^{-\frac{1}{4} \bar{r}^{\rm T} \Omega^{\rm T} \sigma_0 \Omega \bar{r}  +  i  \bar{r}^{\rm T} \Omega^{\rm T} r_0}  \hat{\mathcal{D}}_{\bar{r}} \otimes \hat{\varrho}^q_0   \;, 
\end{equation}
where, for both NPs in the ground state, the initial first and second moments (also known as covariance matrix) are $r_0 = (0,\; 0, \; 0, \;0)^{\rm T}$  and $\sigma_0 = \mathds{1}_4$, respectively.

The Hamiltonian of Eq.~(\ref{eq:hamiltonian_unitary}) is quadratic in the CV operators, where the Gaussain interaction induces a frequency shift and a $g x_1 x_2$ interaction term (containing two-mode squeezing)~\cite{serafini_quantum_2017,ivan_twoqubit_2024,weiss_large_2021} and has linear and diagonal coupling between the qubits and the CV. Thus, the dynamics is integrable~\cite{braccini_superposition_2025}: under this type of Hamiltonians an initial Gaussian state evolves to a Gaussian-branched cat state, which can be similarly described as 
\begin{equation}
\label{eq:density_phase_space}
    \hat{\varrho} = \frac{1}{(2 \pi)^4} \sum_{j, k, m,n \in \pm 1}  \int_{\mathds{R}^{4}} \text{d} \bar{r} \; \chi_{jk}^{mn} (\bar{r})  \hat{\mathcal{D}}_{\bar{r}} \otimes \ket{j} \bra{k} \otimes \ket{m} \bra{n} \;, \nonumber
\end{equation}
where $\ket{j}$ are the elements of the qubits' computational basis and the branched characteristic functions ($\chi_{jk}^{mn}$) are  
\begin{equation}
\label{eq:chara_jk}
   \hspace{-0.16cm} \chi_{jk}^{mn} (\bar{r}) = \left(\varrho^{q} \right)_{jk}^{mn} \exp \left( -\frac{1}{4} \bar{r}^{\rm T} \Omega^{\rm T} \sigma \Omega \bar{r}  +  i  \bar{r}^{\rm T} \Omega^{\rm T} r_{jk}^{mn}  \right) ,
\end{equation}
i.e. Gaussians labeled by the qubits eigenvalues $j,k,m,n$. 

The time-dependent (often complex) parameters $ \sigma$, $r_{jk}^{mn}$, and $\left(\varrho^{q} \right)_{jk}^{mn}$ fully characterise these states and are the covariance matrix, the vectors of the first moments, and the elements of the Qubit Reduced Density Matrix (QRDM), respectively.  This state parameterisation allows us to treat unitary and open dynamics as superposition of quantum Gaussian processes, by mapping the time evolution of the infinite dimensional $\hat{\varrho}(\tau)$ to the evolution of the state's parameters. In the case of unitary dynamics and linear coupling, this is performed via a single time-dependent four-dimensional symplectic transformation induced by the quadratic term of the Hamiltonian~\cite{braccini_superposition_2025}. As shown in Appendix~\ref{app:entangling}, this term can always be rewritten as $\hat{H}_2 = \hat{r}^{\rm T} H_g \hat{r}/2$, where $ H_g$ is a four-dimensional matrix, such that the family of symplectic transformations parameterized by $g$ is \mbox{$S_g(\tau) = \exp( \tau \Omega H_g)$}, with $S_g^{\rm T} \Omega S_g = \Omega \; \forall \tau$.

The covariance matrix $\sigma$ captures the spread and uncertainty of all branches of the wavefunction as it is unaffected by linear terms of the Hamiltonian: as the time evolution of the covariance matrix can be computed as $\sigma(\tau) = S_g(\tau) \sigma_0 S_g(\tau)^{\rm T}$, leading to entangling correlations of two Gaussian states that interact through a \mbox{$g x_1 x_2$~\cite{qvarfort2020mesoscopic,fujita_inverted_2023}}. 
%, as $\sigma = \text{Tr}_m [ \{ (\hat{r}-r), (\hat{r}-r)^{\rm T} \} \hat{\varrho}_{jk}^{mn} ] $, \textcolor{red}{j,k?} where $\{ \cdot, \cdot\}$ represents anti-commutation.  
For example, the element \mbox{$\sigma^{00} = \braket{\hat{x}^2} - \braket{\hat{x}}^2$} encodes the position spread. The Heisenberg uncertainty principle reads $\sigma - i \Omega \geq 0 $~\cite{serafini_quantum_2017}.

The 16 4-dimensional vectors of the first moments $r_{jk}^{mn}$ encode the CV expectation values of each branch of the density matrix. %% $r_{jk}^{mn} := \text{Tr}_m [ \hat{r} \hat{\varrho}_{jk}^{mn} ]$. While,  and are complex on the off diagonal. 
While they can be complex, on the diagonal elements, they are real expectation values of a quantum Gaussian state. In fact, for $j=k$ and $m=n$, the branched characteristic functions are those of Gaussian states, which reveals the statistical mixture of Gaussian processes encoded in the diagonal terms of the density matrix. In this specific case, the time-evolving $r_{jj}^{mm}(\tau)$ represent the four possible interferometric paths taken by the NPs and are plotted in Fig.~\ref{fig:diagram}. At time $\tau = 2 \pi$, i.e. when a single non-interacting SGI recombines~\cite{scala_matter_2013}, the wavefunctions associated with the states $\ket{00}$ and $\ket{11}$ recombine, while a small superposition is still present between the states $\ket{01}$ and $\ket{10}$. One may argue that this effect is caused by the change in frequency induced by the potential. However, by defining the new unitless frequency $\omega_g =  \sqrt{1 - 2g}$ and the final time $\tau_f = 2 \pi/\omega_g$, this effect is still present: the states associated with $\ket{01}$ and $\ket{10}$ recombine, but the superposition between $\ket{00}$ and $\ket{11}$ is  $\delta x_g =  4 f_q \sin^2 (\pi/\omega_g ) $. 
Thus, \textit{the action of sensing entanglement with SGIs induces an operator-valued deflection that does not allow for a perfect recombination of all interferometric paths after an interferometric cycle.}

\begin{figure}[!t]
\includegraphics[width=0.5\textwidth]{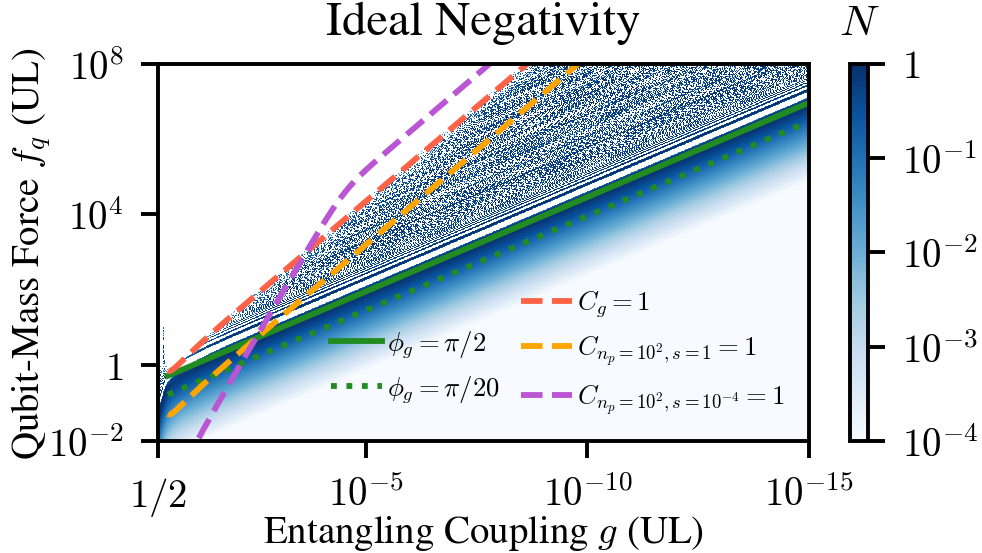}
  \caption{\label{fig:nega} Negativity as function of the unitless parameter space for unitary dynamics with ground-states.}% The common shift of the centers of the SG interferometers has been already considered.}
\end{figure}

We consider entanglement detection via measurements of the qubits: all the information about measurements' outcomes -- and hence entanglement correlations -- is encoded in the QRDM, representing the central result of this work. From the orthonormality of displacement operators, the QRDM is readily computed from Eq.~(\ref{eq:chara_jk}) by noticing $\hat{\varrho}^{q} := \text{Tr}_m [\hat{\varrho}] = \hat{\chi} (0) $, and its components can be written as 
\begin{equation}
     \varrho^q (\tau)_{jk}^{mn} = \left(\varrho_{0}^{q} \right)_{jk}^{mn} \exp \left(- \mathcal{C}_{jk}^{mn} (\tau) + i  \phi_{jk}^{mn}(\tau)  \right)\;, 
\end{equation}
where $\left(\varrho_{0}^{q} \right)_{jk}^{mn} $  are the components of the initial QRDM, and $\phi_{jk}^{mn}(\tau) $ and $\mathcal{C}_{jk}^{mn}(\tau) $ are both time-dependent real quantities and capture the phases that one aims to sense and the decays in coherence encoding noise accumulated during the dynamics (for instance, they are functions of $\sigma$, see Appendix~\ref{app:entangling}). The Positive Partial Transpose (PPT) criterion is a necessary and sufficient condition for entanglement in two-qubits systems~\cite{horodecki_separability_1996,horodecki_quantum_2009}: the (normalized) negativity $\mathcal{N} = - 2 \lambda^{\text{PT}}$ (where $\lambda^{\text{PT}}$ is negative eigenvalue in the partially transposed density matrix) detects entanglement  $1\geq \mathcal{N}>0$. In Appendix~\ref{app:nega}, it is analytically computed alongside the corresponding witness operator $\hat{\mathcal{N}} := - 2 \left(\ket{\lambda^{\text{PT}}} \bra{\lambda^{\text{PT}}}\right)^{\text{PT}}$  -- i.e. the partial transposition of the outer product of the eigenvector $\ket{\lambda^{\text{PT}}}$ corresponds to the negative eigenvalue. The operator $\hat{\mathcal{N}}$ is well approximated (for small $g$) by the witness operator $\hat{\mathcal{W}} =  \frac{1}{2} \left(\hat{\sigma}_x \otimes \hat{\sigma}_x + \hat{\sigma}_y \otimes \hat{\sigma}_z + \hat{\sigma}_z \otimes \hat{\sigma}_y   -  \mathds{1} \otimes \mathds{1} \right)$~\cite{chevalier_witnessing_2020}, which will be considered in the following. $\hat{\mathcal{W}}$ is experimentally realizable via phase estimation measurements and local rotations of the qubits. These measurements have highest visibility when the CV wavefunction is maximally recombined: we consider the final time $\tau_f = 2 \pi/\omega_g$, but it should be noted that, at leading orders in $g$, all the following results are equivalent to the case of $\tau = 2 \pi$.  \mbox{At $\tau_f$, the normalised negativity detectable via $\hat{\mathcal{W}}$ is} 
\begin{equation}
\label{eq:nega}
  \mathcal{N} \approx \text{Tr}_q [ \hat{\mathcal{W}} \hat{\varrho}_q^f ] =  {\rm e}^{-\mathcal{C}_g} \sin(\phi_g)  - \frac{1}{4} \left(1 - {\rm e}^{-4\mathcal{C}_g} \right) \;,
\end{equation}
where the entangling phase is 
\begin{equation}
\label{eq:entanglement_phase}
    \phi_g = \frac{4 \pi g f_q^2}{\omega_g^3} + f_q^2   \sin \left(\frac{2 \pi}{\omega_g}\right) \;,
\end{equation}
which encodes the non-classical correlation accumulated over the dynamics, and the decay of coherence
\begin{equation}
\label{eq:entanglement_contrast}
    \mathcal{C}_g =  2 f_q^2  \sin ^2\left(\pi / \omega_g\right)  \;,
\end{equation} 
which is the dynamical decoherence induced on the qubit due to the remaining superposition at the end of the interferometry. The final QRDM parameters ($\phi_{jk}^{mn} (\tau_f)$ and $\mathcal{C}_{jk}^{mn}(\tau_f)$) are multiples of $\phi_g$ and $\mathcal{C}_g$. In Fig.~\ref{fig:nega}, negativity is given as a function of $f_q$ and $g$ for the ideal case. 

%\subsection{Ideal Unitary Dynamics}

 %, we give the explicit form of $S_g(\tau)$ induced by the Hamiltonian of Eq.~(\ref{eq:hamiltonian_unitary}), and subsequently the time evolution of $\sigma$, $r_{jk}^{mn}$, $\mathcal{C}_{jk}^{mn}$ and $\phi_{jk}^{mn}$, which analytically solves the unitary dynamics of two entangling Stern-Gerlach interferometers with initial general Gaussian state interacting via a $g \hat{x}_1 \hat{x}_2$ potential.

This analytical solution allows us to give order of magnitude estimates useful for experiments: by expanding to the leading order in the small parameter $g$, the phases and contrasts are $\phi_g \sim  6 \pi f_q^2  g$ and $\mathcal{C}_g \sim 2 \pi^2 f_q^2 g^2  $, being a first and second order effects, respectively. Moving back to unitful quantities, they read 
\begin{equation}
\label{eq:phase_unitful}
     \phi_g \sim \frac{6 \pi G F_q^2}{\hbar d^3 \omega^5},  \hspace{0.5cm} \mathcal{C}_g \sim \frac{2 \pi^2 G^2 F_q^2 M}{\hbar d^6 \omega^7} \;.
\end{equation}
For frequencies and forces that are mass-independent, the entangling phase is \textit{mass-independent}, while the decay induced by the unwanted deflection is linear in $M$. It should be noted that, also in the case of Casimir interaction, the entangling phase is mass-independent (it is a function of the density of the NPs), while, for Coulomb, the phase scales as $M^{-2}$ (see Appendix~\ref{app:entangling}).

%along with local phases for each single qubit state given by the inclusion of the force in the interferometers $\phi_{f_g} = 4 \pi f_g f_q$, as in the case of sensing an unknown force with one SG interferometer [], along with the phases from the qubits oscillations, $\phi_0 = 2 \pi \omega_q/\omega$. 

%\subsection{Entanglement}

 %Given that $f(\mathcal{C}_g)>0$ for $\mathcal{C}_g>0$, it follows that $ \mathcal{N} < \frac{1}{2} | \sin (\phi_g) | $, which represents the deviation from the ideal case without fundamental HD effects ($\mathcal{C}_g \to 0$), and is maximised for $\phi_g = \pi$.  

This result allows us to reduce the parameter space of the experiment by imposing a constrain on the phase. We define the detectable phase as $\phi_g^R   :=  6 \pi g (f_q^{\text{R}})^2   = \pi/20$, where $ f^{\text{R}}_q $ is the required force needed to sense entanglement for given coupling~$g$, i.e. $ f^{\text{R}}_q  = 1/\sqrt{120 g}  $ (green lines of Fig.~\ref{fig:nega}). The ideal negativity is $\mathcal{N}_I = \sin(\phi_g) \sim 0.15$ (imposing $\mathcal{C}_g= 0$). This analysis (or any other constraint on the phase) reduces the two parameter space $(f_q, g)$ to only one free parameter (in the following taken to be $g$), or, the four experimental parameters $(F_q, \omega,d,M)$ to three, by setting the mass-independent GIE detection constraint $ F^{\text{R}}_q = \sqrt{ \hbar d^3 \omega^5/(120 G)} $.

%By Taylor expanding in $g$, one finds that $\mathcal{N}\sim  \phi_g - \mathcal{C}_g$. Thus, we define a detectable entanglement when $\phi \sim 6 \pi g f_q^2  \sim \pi/20$ and $\mathcal{C}_g \sim  2 \pi^2 f_q^2 g^2  \ll 1$, such that $\mathcal{N}\sim 0.15$.
%Substituting the former into the latter, it follows that strong qubit-mass coupling is a requirement of entanglement sensing with non-Gaussian interferometry, as $f^{\text{R}}_q \gg \pi/(4 \sqrt{2})$.

%\subsection{State Preparation: \\ Initial squeezed thermal states}
\vspace{0.15cm}

\section{Atoms or LIGO Mirrors?} Already at the level of ideal unitary dynamics, the mass-independent result has to be placed in the context of two bounds of the entangling coupling, and hence the mass, as $g$ is linear in $M$. 

A lower bound can be derived by ensuring that the presented Gaussian interaction treatment still holds. In fact, the Gaussian expansion of the potential of Eq.~(\ref{eq:Newton}) holds only for small superpositions (inversely proportional to the $M$), when the non-Gaussian interaction can be omitted: in this case, a fourth order term given by $\hat{H}_4/(\hbar \omega) =  - 3 x_0^2 g/(2 d^2) (\hat{x}_1 - \hat{x}_2)^4$. Let us estimate the energy scale of this effect in comparison to the Gaussian term $\hat{H}_2/(\hbar \omega) = g (\hat{x}_1 - \hat{x}_2)^2$. At the level of expectation values, which can be computed via repetitive derivation of characteristic functions with respect of the phase space variables~\cite{braccini_superposition_2025}, one finds that  $\braket{\hat{H}_4}/\braket{\hat{H}_2} \sim x_0^2/(5 d^2 g)$, which was computed at the maximum superposition ($\tau = \tau_f/2$) under the detection constraint $f_R$. By imposing an order of magnitude difference between the energy scales, the first bound can be derived: $g>g_{\text{min}} = 2 (x_0/d)^2$. 

An upper bound can be derived by ensuring that the fundamental deflection does not spoil the entanglement: this additional requirement $\mathcal{C}_g  \sim 2 \pi^2 f_q^2 g^2 \ll 1$, implies that $f_q \ll 1/(\sqrt{2} \pi g)$ (red line of Fig.~\ref{fig:nega}). By imposing the detection constraint (and expanding Eq.~\ref{eq:nega} for small~$C_g$), this effect is negligible for $g  \lesssim \frac{60 \mathcal{N}_I}{\pi^2 (1+ \mathcal{N}_I)}\sim 0.8$, which provides a limit of the maximum detectable entanglement coupling. Since from stability of the trap $g<1/2$  (as it is required in order to ensure $\omega_g = \sqrt{1 -2g} \in \mathds{R}$), it is then possible to conclude that the unwanted deflection is not a constraint in the ideal case (and we shall set $g_\text{max} = 1/2$). Thus, combining the two, the mass is bounded between 
\begin{equation}
\label{eq:mass_bound_unitary}
   \sqrt{\frac{\hbar d \omega}{G}} \lesssim  M \lesssim  \frac{d^3 \omega^2}{2 G}
\end{equation}
which for distance of $d \sim 30 \mu$m and $\omega \sim 0.1$Hz reads
$ 10^{-15} \lesssim  M \lesssim 10^{-6}$ kg, spanning the mesoscopic mass scale, which includes neither LIGO mirrors nor atoms. 

\section{State Preparation: \\
Initial Squeezed Thermal States}  

When considering a realistic experimental implementation, these bounds are tightened by the inclusion of noise. The $1/\omega^5$ dependency of the entangling phase (Eq.~\ref{eq:phase_unitful}) highlights the importance of a low-frequency experiment for entanglement detection. However, high frequency traps are experimentally more suitable for trapping and cooling of NP centres of mass (for instance, via optical tweezers with frequencies $\omega_t \sim 100$kHz~\cite{ballestero_levitodynamics_2021}). We shall consider an imperfect cooling in a high-frequency trap followed by the SG dynamics in a ``dark" trap of lower frequency ($\omega \sim 1-100$Hz). Our detection constraint allows us to give a clear account to whether this state preparation -- which includes both squeezing induced by the frequency change~\cite{jansky_strong_1992} and remaining thermal phonons in the CVs initial state -- affects the entangling dynamics.

\begin{figure}[!t]
\includegraphics[width=0.45\textwidth]{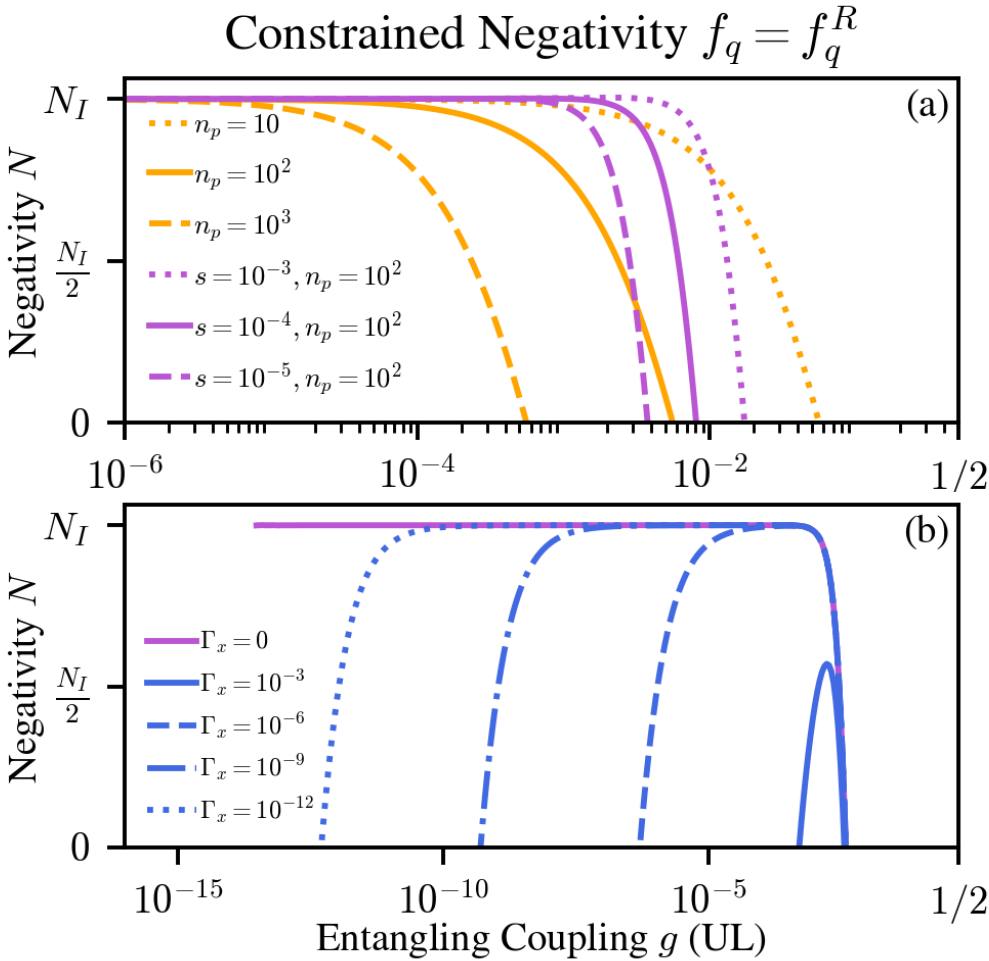}
  \caption{\label{fig:nega_contraint} Constraint negativity  (a) for different initial states and (b) for a squeezed thermal state ($n_p = 10^{2},\;s = 10^{-4}$) undergoing diffusive dynamics as function of the entangling coupling $g$.}% The common shift of the centers of the SG interferometers has been already considered.}
\end{figure}

The decay in contrast induced by the fundamental deflection is amplified when considering squeezed thermal states. The presented formalism allows an immediate treatment by replacing the initial covariance matrix (Eq.~\ref{eq:initial_state}) to $\sigma_0 = (1+2n_p)\; \text{diag} (s, 1/s, s, 1/s)$, where the squeezing parameter, assumed to be equal for both masses, is $s= \omega/\omega_t \sim 10^{-3}-10^{-5}$ and the initial phonons number is $n_p~\sim 0 -100$ -- such that $(1+2n_p) = \coth({\hbar} \omega_t/ (2 k_B T_m))$, being the Boltzmann distribution, where $k_B$ is the Boltzmann constant and $T_m$ the temperature of the CV. As shown in Appendix~\ref{app:open}, with an initial squeezed thermal state for the two NPs, the entangling phase remains \textit{unchanged}, while the decay in contrast at $\tau_f$ is given by
\begin{equation}
 \mathcal{C}_{s,n_p} = (1+2 n_p) \frac{\mathcal{C}_g}{2} \left(\left(s-\frac{1}{s}\right) \cos \left(\frac{2 \pi }{\omega_g}\right)+s+\frac{1}{s}\right) \;. \nonumber
\end{equation}
Given that the structure of the QRDM is the same, the negativity is given by Eq.~(\ref{eq:nega}), under the replacement $\mathcal{C}_g \to \mathcal{C}_{s,n_p}$. Thus, squeezed thermal states only affect the negativity through the fundamental deflection as a second-order effect in $g$: $\mathcal{C}_{s,n_p} = 1$ are given in yellow and purple in Fig.~\ref{fig:nega} for thermal states and thermal-squeezed states, respectively (see Appendix~\ref{app:plots} for supplementary plots). As shown in Fig.~\ref{fig:nega_contraint}, when imposing the detection constraint, the state-preparation only changes the upper bound of $g$ (and hence of Eq.~\ref{eq:mass_bound_unitary}). For thermal state, the bound on the maximum detectable $g$ decreases to  $g_{\text{max}}^{n_p} \lesssim 0.8/(1+2 n_p)$. For highly squeezed states (as the presented case, where $s > (g_{\text{max}})^3$), one has to consider higher order terms in $g$ (see purple line of  Fig.~\ref{fig:nega}) to find that $g_{\text{max}}^{s,n_p} \lesssim  0.45 \left( s/(1+2n_p) \right)^{1/3} $. %\left( \frac{60 s \mathcal{N}_I}{\pi^4 (1+2n_p) (1+ \mathcal{N}_I)} \right)^{1/3}\sim  .

%, which is negligible for small coupling being a second order effect and easily experimentally accessible number of phonons ($n_p \sim 100$). The effect of squeezing is fully captured from the form of $\mathcal{C}_{s,n_p}$ (and the resulting behavior, see purple line Fig~\ref{fig:nega}) which shows a leading order $\mathcal{C}_{s,n_p} 2 (1+2 n_p) \pi s f_q^2 g^2$ for small couplings and squeezing, while, when imposing the detection constraint under high squeezing, one has to consider higher order terms $\mathcal{O}(g^3/s)$.

%It is then possible to conclude that the number of phonons and the squeezing parameter in the initial state are a second order effect in the entangling coupling~$g$: by imposing $\mathcal{C}_{s,n_p} \sim  s (1+2 n_p) \mathcal{C}_g  \ll 1$ and $f_q = f^{\text{R}}_q $, thermal effects only decreases the maximum detectable entanglement given by.  it follows that state preparation effects are negligible for $f_q \ll 1/(\sqrt{2 s (1+2 n_p)} \pi g)$ (purple lines of Fig.\ref{fig:nega}). Strickling, lower entanglement couplings require less cooling of the centre of mass degrees of freedom: for any gravitational experiment, with $s\sim 10^{-4}$ and $ $ by imposing

%\section{Open Dynamics}

\vspace{0.15cm}

\section{Open Dynamics}

A set of leading-order experimental noises, arising from the coupling between the system and the environment, are captured by the master equation
\begin{align}
    \label{eq:master}
    \hspace{-0.2cm}
    \frac{\partial \hat{\varrho}}{\partial \tau} =  \frac{i}{\hbar \omega} \left[ \hat{\varrho} ,\hat{H}_g \right]   - \sum_{i \in 1,2}  \Gamma_x \left[ \hat{x}_i, \left[\hat{x}_i , \hat{\varrho} \right] \right] + \frac{\Gamma_z}{2 \pi} \left[ \hat{\sigma}_z^{(i)}, \left[\hat{\sigma}_z^{(i)} , \hat{\varrho} \right] \right] \nonumber, 
\end{align}
where $\hat{H}_g$ is the Hamiltonian of Eq.~(\ref{eq:hamiltonian_unitary}), the second term captures the diffusion processes in the CVs of rate~$\Gamma_x$, and the last term is the dephasing of each qubit at a rate~$\Gamma_z$. The former Lindbladian term arises from stochastic forces of power spectrum $ S_{\text{FF}}$ (which can be function of $M$, $\omega$ and $d$, depending on the source of noise), such that $\Gamma_x = \pi S_{\text{FF}}/(\hbar M \omega^2)$~\cite{weiss_large_2021}, describing many physical processes, such as scattering of thermal photons or seismic noise in the long-wavelength limit~\cite{schlosshauer_quantum_2007,toros_relative_2021,schut_dephasing_2024}. %Accumulation of random phases by the qubits implies that $\Gamma_z = \pi S_{\text{FF}}(\omega)/(\hbar M \omega^2)$. 
Under this source of noise, the dynamics is still analytically solvable and the entangling phase is found to be unchanged (see Appendix~\ref{app:open}). By performing a similar analysis, entanglement is detectable for 
$ C_{s,n_p} + C_{x} +  C_{z} < \mathcal{N}_I/(1+ \mathcal{N}_I) $, where $C_{x} \approx 3 \pi \Gamma_x f_q^2$ and $ C_{z} \approx 2 \pi \Gamma_z $. Acceptable dephasing implies that $\Gamma_z \lesssim 1$, where we recall that $\Gamma_z$ is in units of $\omega$.  By imposing the detection constraint, the diffusion process bounds the minimum detectable entanglement coupling to $g_{\text{min}} \sim \pi \Gamma_x (1+\mathcal{N}_I)/(40 \mathcal{N}_I) \sim \Gamma_x/2$ (see Fig~\ref{fig:nega_contraint}). \textit{Strong coupling is a requirement for SGIs entanglement detection}. Thus, using the two more-stringent bounds including noises, one finds that entanglement is detectable with masses
\begin{equation}
    \sqrt{\frac{\pi d^3 S_{\text{FF}}}{2 G \hbar}}\lesssim M \lesssim  \left(\frac{s}{1+2 n_p} \right)^{1/3} \frac{d^3 \omega^2}{2 G}\; , \nonumber
\end{equation}
which for $d \sim 30 \mu$m, $\omega \sim 0.1$Hz, $\sqrt{S_{\text{FF}}} \sim 10^{-32}\text{N}/\sqrt{\text{Hz}}$,  $s = 10^{-4}$, and $n_p = 10$, contains the mass to \mbox{$M\sim 10^{-9}$Kg}.

\section{NV-Centres as Qubits} 

When considered a potential experimental implementation, the qubit force and frequency are related to physical quantities. In the case a nanodiamond with an embedded NV centre, the coupling to the qubit is induced by the magnetic gradient ($ \partial B$) that unavoidably traps the NP via its diamagnetic properties~\cite{scala_matter_2013,pedernales_motional_2020,marshman_constructing_2022,shafaq_diamagnetic_2025}. Specifically, $ 
    \omega = \sqrt{|\chi_m|/\mu_0} \partial B  $
    and $
    F_{q} = \tilde{g} \mu_B \partial B$
(where $\tilde{g}$ is the g-factor, $ \mu_B$ the Bohr magneton, $\mu_0$ the vacuum permeability, $ \chi_m$ magnetic susceptibility). Being both linear in $\partial B$, the experimental parameter space reduces to $(\partial B, M, d)$ and the entangling phase of Eq.~(\ref{eq:phase_unitful}) is $\phi_g \sim 10^{-15} (\partial B d)^{-3} \text{T}^3$ (Tesla), which can be used to relate $d$ to~$\omega$: the GIE detection constraint reads  $\omega d \approx 1.6 \cdot 10^{-6} $m/s. The minimum distance -- for which the  unwanted Casimir interaction can be shielded \cite{van_de_kamp_quantum_2020,schut_relaxation_2023,schut_micronsize_2023} -- is $d \sim 30 \mu$m, thus, fixing the maximum experimental frequency to be $\omega_{NV}= 0.05$Hz. This captures the necessity of either a stronger coupling between the qubits and the mass, or an expansion protocol to further delocalize the initial superposition. It should be noted that, for the latter, we show that, exploring an unstable Gaussian protocol, GIE remains mass-independent~\cite{braccini_exponential_2024}. 

\section{Conclusions} 

We provide an analytical treatment of two non-Gaussian matter-wave interferometers that entangle through the leading order Gaussian interaction. Using SGIs, entanglement can be detected via qubit measurements and we show that the leading order entangling phase is independent of the mass of the NPs. In the unitary case, unwanted higher-order terms in the interaction and a fundamental deflection bound the mass from below and above, respectively, to the mesoscopic regime. When considering experimental noise, the analytical solution of the diffusive-dephasing dynamics with initial squeezed thermal state (arising from state preparation) can be formulated in terms of tighter bounds, demanding high isolation and NPs coherence, and ultimately restricting the experiment to Planck mass scales.  

\acknowledgements L.B. would like to acknowledge Engineering and Physical Sciences Research Council (EPSRC) grants (EP/R513143/1 and EP/W524335/1). S.B. would like to acknowledge EPSRC grants (EP/N031105/1, EP/S000267/1, and EP/X009467/1) and grant ST/W006227/1.
This work was made possible through the
support of the WOST, WithOut SpaceTime project (https://withoutspacetime.org), supported by Grant ID\#63683
from the John Templeton Foundation (JTF). S.B’s research is funded by the Gordon and Betty
Moore Foundation through Grant GBMF12328, DOI
10.37807/GBMF12328, and the Alfred P. Sloan Foundation under Grant No. G-2023-21130

\onecolumngrid

\appendix

\section{Intuitive Understanding of Mass Independence from Statical Superpositions\label{app:consideration}}

According to the original argument of Ref.~\cite{bose_spin_2017}, two static superposed masses (with initial state $\ket{\psi}_i = \frac{1}{\sqrt{2}}\left( \ket{L}_i + \ket{R}_i \right)$, where the ``Left" and ``Right" state are eigenstates of $X_i$ and $i\in1,2$ labeling the masses), interacting via a Newtonian potential  
\begin{equation}
    \ket{\psi} = \frac{1}{2} \left[\ket{L}_1 \left( \ket{L}_2 + e^{i \Delta \phi_{LR}} \ket{R}_2 \right) + \ket{R}_1 \left( e^{i \Delta \phi_{RL}} \ket{L}_2 +  \ket{R}_2 \right) \right]
\end{equation}
The entanglement is related to the sum of the relative phases, such that
\begin{equation}
    \Delta \phi_{LR} + \Delta \phi_{RL}  = \frac{G M^2 \tau}{\hbar d} \left( \frac{1}{1 - \Delta x/d} + \frac{1}{1 + \Delta x/d} - 2 \right) \approx \frac{2 G M^2  \Delta x^2}{\hbar d^3} \tau 
\end{equation}
where we Taylor expanded to second order in the small parameter $\Delta x / d$. By noticing that for two SGI, the superposition is given by~\cite{scala_matter_2013} 
\begin{equation}
    \Delta x = \frac{2 \sqrt{2} F_q}{M \omega^2} \implies \;\;\;\;\Delta \phi_{LR} + \Delta \phi_{RL}    = \frac{16  G F_q^2}{\hbar d^3 \omega^4} \tau\;\;\;\;\;\; 
\end{equation}
which shows that the rate of entanglement generation is indeed independent of the mass. At the final time of a SGI $\tau = 2 \pi/ \omega$, when qubit measurements can be performed, the phase scales as in Eq.~(\ref{eq:phase_unitful}).

\section{Entanglement Coupling for Arbitrary Rotation \label{app:entanglement_coupling}}

Gaussian expansion of Newtonian potential is a procedure extensively performed in literature, for example, in GIE entanglement experiment~\cite{weiss_large_2021}, where an arbitrary potential was considered, and in ion trap literature, where a general orientation of the Coulomb interaction is discussed~\cite{ivan_twoqubit_2024}. Here, we consider a general interaction with a general orientation, including higher order terms from the expansions. Two identical particles of equal mass $M$ are trapped at the same frequency $\omega$ . The position operators $\hat{X}_i$, with $i \in \{1,2\}$, of the $i$-th mass measure the displacement in the $x$-direction from its trap centre, and the displacements in the other axes will be omitted. The centres of the two traps are taken to be at $\mathbf{r}_1 = (0,0,0)^{\rm T}$ and $\mathbf{r}_2 = (d \cos\theta, d \sin \theta, 0)^{\rm T}$, where $\theta$ is the angle between $\mathbf{r}_2$ and the $x$ axis, and $| \mathbf{r}_1 - \mathbf{r}_2| = d$ is the distance. In order to study the coupling strength of, let us consider the general potential   
\begin{equation}
\label{eq:general_potential}
    \hat{V} = - \frac{A}{ |\hat{\mathbf{d}}|^{n}}
\end{equation}
where $A$ the coupling strength and $n$ the power of interaction. The operator-valued distance between two masses $\hat{\mathbf{d}}$ with arbitrary orientation is 
\begin{equation}
    \hat{\mathbf{d}} =( (\hat{X}_1 - \hat{X}_2) -  d \cos\theta, \;\; - d \sin \theta, \;\;  0)^{\rm T} \nonumber
\end{equation}
By computing the norm $\left| \hat{\mathbf{d}} \right|$, the potential of Eq.~(\ref{eq:general_potential}) reads
\begin{align}
    \hat{V}_G &= - \frac{A}{d^n \left[1 + \left(\frac{\hat{X}_1 - \hat{X}_2}{d} \right)^2 - 2 \cos \theta \left( \frac{\hat{X}_1 - \hat{X}_2}{d} \right) \right]^{n/2}} \nonumber
\end{align}
Moving to the unit-less operator $\hat{X}_i = \sqrt{2} x_0 \hat{x}_i$, and Taylor expanding for small displacements for large distances compared to the ground state spread \textit{i.e.} $d \gg x_0$,  such that 
\begin{align}
    \frac{2 \hat{V}_G}{\hbar \omega}  = co&nst -  f(\theta) \left( \hat{x}_1 - \hat{x}_2 \right)   - g(\theta) \left( \hat{x}_1 - \hat{x}_2 \right)^2 - h(\theta) \left( \hat{x}_1 - \hat{x}_2 \right)^3 - p(\theta) \left( \hat{x}_1 - \hat{x}_2 \right)^4 \;,
\end{align}
where the constant will be omitted and the first two terms 
\begin{equation}
    f (\theta) = 2 \sqrt{2} n \frac{ A  x_0}{\hbar \omega d^{n+1}}  \cos \theta \hspace{1cm}
     g (\theta) =  \frac{ A x_0^2 n}{ \hbar \omega d^{n+2}} \left(n + (n + 2) \cos (2 \theta) \right) 
\end{equation}
where the former representing the mean force and the second is the entangling coupling. The next two higher order terms are given by 
\begin{equation}
    h (\theta) = \frac{2 \sqrt{2}n (n+2)}{3} \frac{ A  x_0^3 }{ \hbar \omega d^{n+3}}  \cos \theta \left( (n+4)  \cos^2 ( \theta) - 3 \right) \; , \nonumber
\end{equation}
\begin{equation}
    p(\theta)  = \frac{n (n+2)}{3} \frac{A  x_0^4  }{ \hbar \omega d^{n+4} }\left((n+4) \cos ^2(\theta ) \left((n+6) \cos ^2(\theta )-6\right)+3\right) \nonumber
\end{equation}

In Table~\ref{tab:ent_coupling}, we give the explicit form of Coulomb, Gravity and Casimir interaction for linear ($\theta = 0$) and parallel ($\theta = \pi/2$) orientations~\cite{schut_improving_2022,braccini_large_2023}. The former has a non-zero mean force which leads to non-entangling local phases between the SGIs, while the later does not. We note that the angular dependency of the entanglement coupling ($g_G(\theta)$) is given by $n + (2 + n ) \cos (2 \theta)$. The strength, measured in absolute value, has its maxima for $\theta = k \pi$ with $k \in \mathds{N}$, i.e. the linear case, while, for $\theta = (k + 1/2) \pi$ with $k \in \mathds{N}$, i.e. parallel case, are local minima of $|g_G(\theta)|$. 

However, we still chose the parallel orientation as in this direction low frequencies are easier to implement in an experimental setting. The absence of the linear force allows for a more compact description of states and evolution. Furthermore, in regards of higher order term, the parallel orientation implies that the cubit term $h(\theta = \pi/2) =0 $ and the forth order term has to be considered, thus making this orientation more robust for the Gaussian treatment.

\begin{table*}
\begin{center}
\captionof{table}{\label{tab:ent_coupling} The
Unit Less (UL) couplings between two levitated masses interacting via Coulomb, Casimir and Newtonian interactions as function of: mass ($M$), frequency of the trap ($\omega$), distance ($d$), additional charge for Coulomb ($Q$), relative permittivity ($\epsilon$), mass density ($\varrho_m$). $c$ is the speed of light, $\hbar$ the Plank constant, $\epsilon_0$ the vacuum permittivity and $G$ the Newton constant.}
\begin{tabular}{ |p{4.5cm}|p{1cm}||p{3.7cm}|p{3.7cm}|p{3.7cm}| }
 \hline
 Parameters & Symbol& Coulomb & Casimir &  Newtonian \\
 \hline
 \hline
 Interaction Strength  ($\text{N}/\text{m}^n$) & A & $- Q^2/(4 \pi \epsilon_0)$ &  $\frac{207}{64 \pi^3}  \left( \frac{\epsilon - 1}{\epsilon +2} \right)^2 c \hbar M^2/\varrho_m^2$ & $G M^2$   \\
 Interaction Power  & n & $1$ & $7$ & $1$ \\
 \hline
 \multicolumn{5}{|c|}{Linear Orientation ($\theta = 0$)} \\
  \hline
  Force &$f_l$  & $ - Q^2/ \left(2 \pi \epsilon_0  \sqrt{\hbar M \omega^3} d^2 \right) $ & $\frac{1449}{32 \pi^3} \left( \frac{\epsilon - 1}{\epsilon +2} \right)^2 \frac{c \sqrt{\hbar }}{d^8 \varrho_m^2} \left( \frac{M}{\omega} \right)^{3/2} $   &  $2 G M^{3/2}/\left( \sqrt{\hbar \omega^3} d^2 \right) $  \\
 Entanglement Coupling  & $g_l$  & $- Q^2/\left( 2 \pi \epsilon_0  M \omega^2 d^3\right)$ &$\frac{1449}{8 \pi^3} \left( \frac{\epsilon - 1}{\epsilon +2} \right)^2 \frac{c \hbar M}{\omega^2 \varrho_m^2 d^9}$    &  $2 G M / \left(d^3 \omega^2\right)$   \\
% Next Order expansion & $h_l$  & $\frac{ Q^2}{4 \pi \epsilon_0 d^3 } \left( \frac{\hbar}{M \omega} \right)^{3/2}$ & Hz   &  10   \\
 \hline
 \multicolumn{5}{|c|}{Parallel Orientation ($\theta = \pi/2$)} \\
  \hline
 Force (UL)&$f_p$  & 0 & 0   &  0   \\
 Entanglement Coupling  &$g_p$  & $-Q^2/ \left( 4 \pi \epsilon_0  M \omega^2 d^3 \right)$ & $\frac{1449}{64 \pi^3} \left( \frac{\epsilon - 1}{\epsilon +2} \right)^2 \frac{c \hbar M}{\omega^2 \varrho_m^2 d^9}$    &  $ G M / \left(d^3 \omega^2\right)$   \\
 \hline
\end{tabular}
\end{center}
\end{table*}

\section{Ideal Entangling Dynamics \label{app:entangling}}

The Hamiltonian  of Eq.~(\ref{eq:hamiltonian_unitary}) can be mapped to the general Gaussian linear-operator-valued Hamiltonian of the form 
\begin{align}
\label{eq:Hamiltonian_general}
    \hspace{-0.3cm}\frac{\hat{H}}{\hbar \omega} &=   \frac{1}{2} \hat{r}^{\rm T} H_m \hat{r} - r_{f}^{\rm T} \hat{r}  - \sum_{i \in \{1,2\}}  \left( r_{q}^{(i)} \right)^{\rm T} \hat{r}  \otimes \hat{\sigma}_z^{(i)}   \;, 
\end{align}
via
\begin{equation}
\label{eq:hamiltonian_matrix}
  H_m = \begin{pmatrix}
       1 - g & 0 & g  & 0 \\
     0 & 1 & 0  & 0 \\
      g & 0 & 1 - g & 0 \\
       0 & 0 & 0 & 1  
  \end{pmatrix} \hspace{1cm}
  r_q^{(1)}= \begin{pmatrix} f_q \\ 0 \\ 0 \\ 0
  \end{pmatrix}
  \hspace{1cm}
  r_q^{(2)}= \begin{pmatrix} 0 \\ 0 \\ f_q \\ 0
  \end{pmatrix}
  \hspace{1cm}
  r_f= \begin{pmatrix} 0 \\ 0 \\ 0 \\ 0
  \end{pmatrix}
\end{equation}
which finds the general solution in terms of the symplectic transformation associated to $H_m$, i.e. $S_{g}(\tau) = {\rm e}^{\tau \Omega H_g} $ and the quantities $r_j^m = j r_q^{(1)} + m r_q^{(2)} + r_m $, from which all the parameters of the GCS solving the dynamics can be derived (Table~\ref{tab:unitary_general}). Defining $\omega_g = \sqrt{1 - 2 g}$, the symplectic transformation solving the unitary dynamics reads
\begin{equation}
    S_{g}(\tau) = 
    \frac{1}{2}  \begin{pmatrix}
        \;\;\;\; \cos \tau   + \cos (\omega_g \tau) 
        \;\;\;\; & \;\;\;\;
        \sin \tau   +  \frac{1}{\omega_g } \sin (\omega_g \tau)   \;\;\;\; & \;\;\;\;
        \cos \tau   - \cos (\omega_g \tau)  
        \;\;\;\; & \;\;\;\; 
        \sin \tau  -   \frac{1}{\omega_g } \sin (\omega_g \tau)  
        \;\;\;\;\\\;\;\;\;
        - \sin \tau   -  \omega_g  \sin (\omega_g \tau)   
        &
        \cos \tau   + \cos (\omega_g \tau)   
        & 
        - \sin \tau   +  \omega_g  \sin (\omega_g \tau)  
        &  
        \cos \tau   - \cos (\omega_g \tau)  
        \\
        \cos \tau   - \cos (\omega_g \tau)  
         & 
        \sin \tau  -   \frac{1}{\omega_g } \sin (\omega_g \tau) 
        & 
        \cos \tau   + \cos (\omega_g \tau) 
        & 
        \sin \tau   +  \frac{1}{\omega_g } \sin (\omega_g \tau) 
        \\
         - \sin \tau   + \omega_g  \sin (\omega_g \tau)  
        &  
        \cos \tau   - \cos (\omega_g \tau)  
         &
         - \sin \tau   -  \omega_g  \sin (\omega_g \tau)
        &
        \cos \tau   + \cos (\omega_g \tau)   
    \end{pmatrix}
\end{equation}

\begin{table*}[!]
    \centering
    \renewcommand{\arraystretch}{1.7}
    \begin{tabular}{ |p{4.0cm}||P{6cm}|P{6cm}|}
    \hline
        Phase Spaces Quantities & \multicolumn{2}{c|}{Solution of Unitary Dynamics with Operator Valued Linear Interactions} \\ 
        \hline
        \hline
        Covariance Matrices   & \multicolumn{2}{c|}{\ensuremath{
   \sigma (\tau)   = S_g (\tau) \sigma_0 S_g^{\rm T} (\tau)}}\\
       First Moments & \multicolumn{2}{c|}{
       \ensuremath{r_{jk}^{mn} (\tau) = r_0(\tau) - \frac{1}{2}  \left( \tilde{r}_j^m(\tau) - \tilde{r}_j^m + \tilde{r}_k^n(\tau) - \tilde{r}_k^n  \right) -  \frac{i}{2} \sigma (\tau)   \Omega  \left[ (\tilde{r}_j^m (\tau) - \tilde{r}_j^m)  - (\tilde{r}_k^n (\tau) - \tilde{r}_k^n ) \right]   }} \\
     QRDM Contrasts & \multicolumn{2}{c|}{
       \ensuremath{\mathcal{C}_{jk}^{mn} (\tau) = \frac{1}{4} \left[ (\tilde{r}_j^m (\tau) - \tilde{r}_j^m)  - (\tilde{r}_k^n (\tau) - \tilde{r}_k^n ) \right]^{\rm T} \Omega^{\rm T} \sigma (\tau)   \Omega  \left[ (\tilde{r}_j^m (\tau) - \tilde{r}_j^m)  - (\tilde{r}_k^n (\tau) - \tilde{r}_k^n ) \right]  }}\\
    QRDM Phases & \multicolumn{2}{c|}{\ensuremath{ \phi_{jk}^{mn} (\tau) = - (\tilde{r}_j^m - \tilde{r}_k^n)^{\rm T} \Omega \left[  ( \tilde{r}_0 (\tau) - \tilde{r}_0) -  \frac{1}{2} \left( \tilde{r}_j^m (\tau) - \tilde{r}_j^m + \tilde{r}_k^n (\tau) -  \tilde{r}_k^n  \right) \right]  }} \\
 & \multicolumn{2}{c|}{\ensuremath{ \hspace{2cm}  + \frac{\tau}{2} \left[ (\tilde{r}_j^m - \tilde{r}_k^n)^{\rm T} H_m (\tilde{r}_j^m + \tilde{r}_k^n) - H_q^0 (j-k)\right] }} \\
    \hline
    \end{tabular}
    \caption{Time evolution of the phase space quantities of a GCS (Eq.~\ref{eq:chara_jk}) undergoing unitary evolution with operator-valued forces (Eq.~\ref{eq:hamiltonian_unitary}), where $\tilde{r}_a = H^{-1}_m r_a$, $S_g (\tau) = \exp (  \Omega H_m \tau)$, and $\tilde{r}_a (\tau) = S_g (\tau) \tilde{r}_a$, with $a \in 0,{}_j^m,{}_k^n$. The initial state is a $n$-modes Gaussian state with first and second moments $r_0$ and $\sigma_0$ and a general QRDM $\varrho^{q}_{jk}(0)$~\cite{braccini_superposition_2025}.}
    \label{tab:unitary_general}
\end{table*}

The joint covariance matrix for two initial uncorrelated ground states (i.e. $\sigma_0 = \mathds{1}_4$) evolves to 
\begin{equation}
\label{eq:evolution_entangled_cov}
\sigma (\tau) =    
    \left(
\begin{array}{cccc}
 \frac{2 - g \left(3 + \cos 2  \omega_g \tau \right)}{ 2 \omega_g^2} 
 & \frac{g }{2  \omega_g}  \sin \left(2  \omega_g \tau \right) 
 & -\frac{g  }{\omega_g^2}  \sin ^2\left( \omega_g \tau \right)
 & -\frac{g }{2  \omega_g}  \sin \left(2  \omega_g \tau \right) 
 \\
 \frac{g }{2  \omega_g}  \sin \left(2  \omega_g \tau \right) 
 & \frac{1}{2} \left(g \cos \left(2  \omega_g \tau \right)-g+2\right) 
 & -\frac{g }{2  \omega_g}  \sin \left(2  \omega_g \tau \right) 
 & g  \sin ^2\left( \omega_g \tau \right) 
 \\
 -\frac{g }{\omega_g^2} \sin ^2\left( \omega_g \tau \right)
 & -\frac{g }{2  \omega_g}  \sin \left(2  \omega_g \tau \right) 
 & \frac{2 - g \left(3 + \cos 2  \omega_g \tau \right)}{ 2 \omega_g^2} 
 & \frac{g}{2  \omega_g} \sin \left(2  \omega_g \tau \right) 
 \\
 -\frac{g }{2  \omega_g}  \sin \left(2  \omega_g \tau \right) 
 & g  \sin ^2\left( \omega_g \tau \right) 
 & \frac{g }{2  \omega_g}  \sin \left(2  \omega_g \tau \right) 
 & \frac{1}{2} \left(g \cos \left(2  \omega_g \tau \right)-g+2\right) \\
\end{array}
\right)
\end{equation}
The first moments on the diagonal term,  plotted in Fig.~\ref{fig:diagram}, are 
\begin{equation}
    r_{00}^{00} (\tau) = - r_{11}^{11}  (\tau) = f_q 
\begin{pmatrix}
 \cos \tau -1 \\
  -\sin \tau  \\
  \cos \tau -1 \\
  -\sin \tau  \\
\end{pmatrix} \hspace{0.5cm}
r_{01}^{01} (\tau) = - r_{10}^{10} (\tau) = \frac{f_q}{\omega_g}
\begin{pmatrix}
  \frac{1}{\omega_g} (\cos \tau  \omega_g - 1) \\
 - \sin (\tau  \omega_g) \\
 \frac{1}{\omega_g} (1 - \cos \tau  \omega_g)\\
 \sin (\tau  \omega_g) \\
\end{pmatrix}
\end{equation}
where, for coincidences, we will omit the off diagonal term which includes complex parts, as given in Table~\ref{tab:unitary_general}. Still, we can compute the time evolution of the QRDM, such that for an initial equal superposition $\ket{+}_{q_1} \otimes \ket{+}_{q_2}$ for the qubit, the QRDM evolves to 
\begin{equation}
    \hat{\varrho}_{q}(\tau) = \frac{1}{4}
\left(
\begin{array}{cccc}
 1 & 
 {\rm e}^{- \mathcal{C}_{1}(\tau) - \mathcal{C}_{2}(\tau) - i \phi_g (\tau)} & 
 {\rm e}^{- \mathcal{C}_{1}(\tau) - \mathcal{C}_{2}(\tau) - i \phi_g(\tau) } & 
 {\rm e}^{- 4 \mathcal{C}_{2}(\tau) }
 \\
  {\rm e}^{- \mathcal{C}_{1}(\tau) - \mathcal{C}_{2}(\tau) + i  \phi_g (\tau)} & 
 1 &
 {\rm e}^{- 4 \mathcal{C}_{1}(\tau) } & 
 {\rm e}^{- \mathcal{C}_{1}(\tau) - \mathcal{C}_{2}(\tau) + i \phi_g(\tau) } \\
 {\rm e}^{- \mathcal{C}_{1}(\tau) - \mathcal{C}_{2}(\tau) + i \phi_g(\tau) }  & 
 {\rm e}^{- 4 \mathcal{C}_{1}(\tau)} & 
 1 & 
 {\rm e}^{-  \mathcal{C}_{1}(\tau) - \mathcal{C}_{2}(\tau) + i \phi_g(\tau) }  \\
 {\rm e}^{- 4 \mathcal{C}_{2}(\tau) } & 
 {\rm e}^{- \mathcal{C}_{1}(\tau) - \mathcal{C}_{2}(\tau) - i \phi_g(\tau) }  & 
 {\rm e}^{- \mathcal{C}_{1}(\tau)  - \mathcal{C}_{2}(\tau)- i \phi_g(\tau) } & 
 1 \\
\end{array}
\right) \nonumber
\end{equation}
where 
\begin{equation}
    \phi_g(\tau) = f_q^2 \left(\sin (\tau ) +  \frac{ 2 g \tau}{ \omega_g^2}  -\frac{\sin ( \omega_g \tau )}{\omega_g^3} \right)
\hspace{0.5cm}
 \mathcal{C}_{1}(\tau) = 
\frac{  2 f_q^2 }{ \omega_g^4} \sin ^2\left(\frac{\tau \omega_g}{2}\right) \left[1 - g\left( 1 + \cos (\tau  \omega_g )\right)\right] \hspace{0.5cm}
\mathcal{C}_{2}(\tau) = 2 f_q^2 (\cos (\tau )-1)\nonumber
\end{equation}
where the former is the entangling phase and the latter two encode the lost of coherence in the qubits state due to the superposition in the CV degree of freedom, known as HD: the former of the two is due the entangling deflection and the latter is the one present also in the single interferometer case. At the time $\tau_f = 2 \pi/\omega_g$, the final QRDM is
\begin{equation}
\label{eq:qrdm_ideal}
    \hat{\varrho}_{q}^f = \frac{1}{4}
\left(
\begin{array}{cccc}
 1 & 
 {\rm e}^{- \mathcal{C}_g - i \phi_g } & 
 {\rm e}^{- \mathcal{C}_g - i \phi_g } & 
{\rm e}^{- 4 \mathcal{C}_g }
 \\
  {\rm e}^{- \mathcal{C}_g + i  \phi_g } & 
 1 &
 1& 
 {\rm e}^{- \mathcal{C}_g + i \phi_g } \\
 {\rm e}^{- \mathcal{C}_g + i \phi_g }  & 
 1  & 
 1 & 
 {\rm e}^{- \mathcal{C}_g + i \phi_g }  \\
 {\rm e}^{- 4 \mathcal{C}_g } & 
 {\rm e}^{- \mathcal{C}_g - i \phi_g }  & 
 {\rm e}^{- \mathcal{C}_g - i \phi_g } & 
 1 \\
\end{array}
\right)
\end{equation}
where the final phase is given by Eq.~(\ref{eq:entanglement_phase}), $\mathcal{C}_{1}(\tau_f) = 0$, and $\mathcal{C}_{2}(\tau_f) = \mathcal{C}_g $, which is given by Eq.~(\ref{eq:entanglement_contrast}). 

\begin{table*}
\begin{center}
\captionof{table}{\label{tab:ent_dynamics} 
Leading order expansion in the unitless coupling $g$ of the entangling phase and contrast, for two Stern-Gerlach interferometers interacting via Coulomb, Casimir and Newtonian interaction as function of Qubit-Mass force ($F_q$),  Mass ($M$), frequency of the trap ($\omega$), distance ($d$), additional charge for Coulomb ($Q$), relative permittivity ($\epsilon$), mass density ($\varrho_m$). $c$ is the speed of light, $\hbar$ the Plank constant, $\epsilon_0$ the vacuum permittivity and $G$ the Newton constant.}
\begin{tabular}{ |p{2.5cm}|p{1cm}||p{3.7cm}|p{3.7cm}|p{3.7cm}| }
 \hline
  Parameters & Symbol& Coulomb & Casimir &  Newtonian \\
 \hline
 \hline
 Entangling Phase 
 & $\phi_g$ 
 &$ - 3  Q^2 F_q^2  /(2 \hbar \epsilon_0    M^2 d^3  \omega^5 )$ 
 & $  \frac{4347}{32 \pi ^2} \left(\frac{\epsilon-1}{\epsilon+2}\right)^2 \frac{ c F_q^2 }{ d^9  \varrho_m^2 \omega ^5}$ 
 & $ 6 \pi G F_q^2/(\hbar d^3 \omega^5)$   
 \\Contrast  & 
 $\mathcal{C}_g$ & 
 $ F_q^2   Q^4 / (8 \hbar \epsilon_0^2 d^6  M^3 \omega ^7 )$ & 
 $\frac{2099601}{2048 \pi^4} \left(\frac{\epsilon-1}{\epsilon+2}\right)^2 \frac{ \hbar c^2 M F_q^2  }{ d^{18}  \varrho_m^4   \omega ^7 }$ & 
 $ 2 \pi ^2 G^2 M  F_q^2 / ( \hbar d^6 \omega ^7)$ \\
 %\\ Detectable Entanglement ? & 
 %$g_D$ & 
 %$ F_q^2   Q^4 / (2 \hbar \epsilon_0^2 d^6  M^3 \omega ^7 )$ & 
 %$\frac{2099601}{32 \pi^4} \left(\frac{\epsilon-1}{\epsilon+2}\right)^2 \frac{ \hbar c^2 M F_q^2  }{ d^6  \varrho_m^4   \omega ^7 }$ & 
 %$ 8 \pi ^2 G^2 M  F_q^2 / ( \hbar d^6 \omega ^7)$ \\
 \hline
\end{tabular}
\end{center}
\end{table*}

\section{Negativity and Approximated Witness\label{app:nega}}

By taking the partial transposition of the QRDM given in Eq.~(\ref{eq:qrdm_ideal}), one finds that the negativity is 
\begin{equation}
  \mathcal{N} = \frac{e^{-C_g}}{2} \left[\sqrt{\sin^2(\phi_g) +  f^2(\mathcal{C}_g)} - f(\mathcal{C}_g) \right] 
\end{equation}
where $f(\mathcal{C}_g) = \frac{1}{2} e^{-\mathcal{C}_g}\sinh (2 \mathcal{C}_g)$. The normalized eigenvector associated to the negative eigenvalue given in Eq.~(\ref{eq:nega}), representing the negativity, is 
\begin{equation}
    \ket{\lambda^{\text{PT}}} = \frac{1}{\sqrt{2+2w^2}} \begin{pmatrix}
      1 \\
       i w \\
      - i w \\
      -1
    \end{pmatrix} \hspace{0.5cm} w = \frac{2 \mathcal{N} {\rm e}^{\mathcal{C}_g}}{\sin(\phi_g)} = \frac{1}{\sin(\phi_g)}  \left(\sqrt{ \sin ^2(\phi_g)+f^2(\mathcal{C}_g)}- f(\mathcal{C}_g) \right)
\end{equation}
from which one can find the witness operator, as the partial transposition of the outer product, 
\begin{equation}
    \hat{\mathcal{N}} = - \ket{\lambda^{\text{PT}}}\bra{\lambda^{\text{PT}}}^{\text{PT}}
    = \frac{- 1}{2+2w^2} 
    \begin{pmatrix}
 1 & i w & i w & -w^2 \\
 -i w & w^2 & -1 & -i w \\
 -i w & -1 & w^2 & -i w \\
 -w^2 & i w & i w & 1 
\end{pmatrix} \;.
\end{equation}
%which, when considering an experimental setting, can be used to find the weighs of the local Pauli measurements that have to be performed on the two qubit systems. 
which is measurable in an experiment through sets of local Pauli measurements. This is due to the fact that $\sigma_j^{(1)} \otimes \sigma_k^{(2)}$, with $j,k \in \{x,y,z\}$, are a complete basis of space of operators acting on the joint 2-qubit Hilbert space. %In situations where the measurement options are limited, alternative 2-qubit entanglement witnesses may be applied~\cite{bose_spin_2017,chevalier_witnessing_2020}. 
In order to simplify the following discussions and reduced the number of required Pauli measurements, we can note that, for small $g$,  $\mathcal{C}_g \to 0$ (being a second order term compared to $\phi_g$) and hence $w \to 1$, such that the witness can be approximated to 
\begin{equation}
\label{eq:witness_approx}
    \hat{\mathcal{N}} \approx \hat{\mathcal{W}} = 
    - \frac{1}{4 }\begin{pmatrix}
 1 & i  & i  & -1 \\
 -i  & 1 & -1 & -i  \\
 -i  & -1 & 1 & -i  \\
 -1 & i  & i  & 1 
\end{pmatrix} 
\end{equation}
which is the witness used in this work and discussed in~\cite{chevalier_witnessing_2020}.

\section{Diffusive Dynamics with Thermal Squeezed State \label{app:open}}
\begin{table*}
    \centering
    \renewcommand{\arraystretch}{1.7}
    \begin{tabular}{ |p{4.0cm}||P{7cm}|P{7cm}|}
       \hline
        Phase Spaces Quantities & \multicolumn{2}{c|}{Operator Valued Linear Interactions (Symmetric Case)} \\ 
        \hline
        Covariance Matrices   & \multicolumn{2}{c|}{\ensuremath{
   \sigma (\tau) := \sigma_{jk}(\tau) = 
    S_m(\tau) \sigma_0 S_m^{\rm T}(\tau) + \int_{0}^{\tau} \text{d}t S_m(\tau - t) D S^{\rm T}_m(\tau - t) }}\\
    \hline
       Vectors or First Moments & \multicolumn{2}{c|}{
       \ensuremath{r_{jk}^{mn} (\tau) = r_0(\tau) - \frac{1}{2}  \left( \tilde{r}_j^m(\tau) - \tilde{r}_j^m + \tilde{r}_k^n(\tau) - \tilde{r}_k^n  \right) - \frac{i}{2}  \sigma (\tau)  \Omega  \left[ (\tilde{r}_j^m (\tau) - \tilde{r}_j^m)  - (\tilde{r}_k^n (\tau) - \tilde{r}_k^n ) \right] }} \\
           & \multicolumn{2}{c|}{
       \ensuremath{ \hspace{2cm}  - \frac{i}{2}  \int_0^{\tau} \text{d}t  D (\tau-t) \Omega  \left[ (\tilde{r}_j^m(\tau - t) - \tilde{r}_j^m(\tau) ) -  (\tilde{r}_k^n(\tau - t) - \tilde{r}_k^n(\tau) ) \right]  }} \\
       \hline
     QRDM Contrasts  & \multicolumn{2}{c|}{ \hspace{-0.1cm} \ensuremath{  \mathcal{C}_{jk}^{mn} (\tau)  =    \frac{1}{4} \left[ (\tilde{r}_j^m (\tau) - \tilde{r}_j^m)  - (\tilde{r}_k^n (\tau) - \tilde{r}_k^n ) \right]^{\rm T} \Omega^{\rm T} \sigma (\tau)   \Omega  \left[ (\tilde{r}_j^m (\tau) - \tilde{r}_j^m)  - (\tilde{r}_k^n (\tau) - \tilde{r}_k^n ) \right]  + \tau \Gamma_z (jk - 1) }} \\
 & \multicolumn{2}{c|}{\ensuremath{ \hspace{1cm}  \frac{1}{4}     \int_0^\tau \text{d} t   \Big\{ \left[ (\tilde{r}_j^m (\tau - t) - \tilde{r}_j^m(\tau))  - (\tilde{r}_k^n (\tau - t) - \tilde{r}_k^n(\tau))  \right]^{\rm T}  \Omega^{\rm T}  D(\tau - t) \Omega  }} \\
 & \multicolumn{2}{c|}{\ensuremath{ \hspace{2.5cm} 
 \left[ (\tilde{r}_j^m (\tau - t)  + \tilde{r}_j^m(\tau) - 2 r_j)  - (\tilde{r}_k^n (\tau - t) + \tilde{r}_k^n(\tau) - 2 r_k)  \right]  \Big\} }} \\
\hline 
    \end{tabular}
    \caption{Open evolution of a GCS phase space quantity under operator-valued linear Hamiltonian (Eq. \ref{eq:hamiltonian_unitary}) and Gaussian Diffusive noise ($D$) of an initial Gaussian state with initial first and second moments $r_0$ and $\sigma_0$, where $\tilde{r}_j^m = H^{-1}_m r_j^m$, $S_m (\tau) = \exp (  \Omega H_m \tau)$, and $\tilde{r}_a (\tau) = S_m (\tau) \tilde{r}_a$, with $a \in 0,{}_j^m,{}_k^n$~\cite{braccini_superposition_2025}.}
    \label{tab:open_general}
\end{table*}

In this Appendix, we give the analytical solution of the entangling dynamics with the inclusion of diffusion on both masses, due to thermal photon scattering. Under the open dynamics of Eq.~(\ref{eq:master}), the phase space quantities parameterizing a general GCS evolves according to Table~\ref{tab:open_general}, where we note that the covariance matrix evolves according to the solution of the Lyapunov equation and the phases are unchanged from Table~\ref{tab:unitary_general}. The initial state considered is a thermal squeezed state -- usually produced by cooling in a tighter trap -- centred in each trap, which implies the first and second moments $r_0 = (0,0,0,0)^{\rm T }$ and $\sigma_0 = (1+2n_p) \text{diag} (s, 1/s, s, 1/s)$, where $n_p$ is the number of phonons and the squeezing parameter $s= \omega/\omega_t$, where $\omega_t$ is the frequency of the cooling trap.

Thus, the covariance matrix at time $\tau$ can be written as $\Sigma (\tau) = (1 + 2n_p) \sigma_s (\tau) + \Gamma_x \tilde{\sigma} (\tau)$, where $\sigma_s$ is the covariance matrix of the unitary dynamics with initial squeezed state and $\tilde{\sigma}$ is induced by the diffusive process. For coincidences, we shall report only the values at $\tau_f = 2\pi/\omega_g$, and, defining the quantity $(\delta s)^2 = 1-s^2$, it is possible to find the analytical form
\begin{equation}
\hspace{-1.2cm}    \sigma_s(\tau_f)    = \frac{1}{4 s \omega_g^2}
\begin{pmatrix}
 \omega_g^2 \left(3 s^2+1 - \delta s^2  \cos \left(\frac{4 \pi }{\omega_g}\right)\right) & \delta s^2 \omega_g^2 \sin \left(\frac{4 \pi }{\omega_g}\right) & 2 \delta s^2 \omega_g^2 \sin ^2\left(\frac{2 \pi }{\omega_g}\right) & \delta s^2 \omega_g^2 \sin \left(\frac{4 \pi }{\omega_g}\right) \\
 \delta s^2 \omega_g^2 \sin \left(\frac{4 \pi }{\omega_g}\right) &  \omega_g^2 \left(3 s^2+1 + \delta s^2  \cos \left(\frac{4 \pi }{\omega_g}\right)\right) & \delta s^2 \omega_g^2 \sin \left(\frac{4 \pi }{\omega_g}\right) & 2 \left(s^2-1\right) \omega_g^2 \sin ^2\left(\frac{2 \pi }{\omega_g}\right) \\
 2 \delta s^2 \omega_g^2 \sin ^2\left(\frac{2 \pi }{\omega_g}\right) & \delta s^2 \omega_g^2 \sin \left(\frac{4 \pi }{\omega_g}\right) & \omega_g^2 \left(3 s^2+1 - \delta s^2  \cos \left(\frac{4 \pi }{\omega_g}\right)\right) & \delta s^2 \omega_g^2 \sin \left(\frac{4 \pi }{\omega_g}\right) \\
 \delta s^2 \omega_g^2 \sin \left(\frac{4 \pi }{\omega_g}\right) & 2 \left(s^2-1\right) \omega_g^2 \sin ^2\left(\frac{2 \pi }{\omega_g}\right) & \delta s^2 \omega_g^2 \sin \left(\frac{4 \pi }{\omega_g}\right) &  \omega_g^2 \left(3 s^2+1 + \delta s^2  \cos \left(\frac{4 \pi }{\omega_g}\right)\right) \\
\end{pmatrix} \nonumber
\end{equation}
and 
\begin{equation}
\tilde{\sigma}(2 \pi)  =
\begin{pmatrix}
 \pi (1 -  g) \omega_g^{2/3} -\frac{1}{8} \sin \left(\frac{4 \pi }{\omega_g}\right) & \frac{1}{4} \sin ^2\left(\frac{2 \pi }{\omega_g}\right) & -\frac{\pi  g}{\omega_g^{3/2}}-\frac{1}{8} \sin \left(\frac{4 \pi }{\omega_g}\right) & \frac{1}{4} \sin ^2\left(\frac{2 \pi }{\omega_g}\right) \\
 \frac{1}{4} \sin ^2\left(\frac{2 \pi }{\omega_g}\right) & \frac{\pi }{\omega_g}+\frac{1}{8} \sin \left(\frac{4 \pi }{\omega_g}\right) & \frac{1}{4} \sin ^2\left(\frac{2 \pi }{\omega_g}\right) & \frac{1}{8} \sin \left(\frac{4 \pi }{\omega_g}\right) \\
 -\frac{\pi  g}{\omega_g^{3/2}}-\frac{1}{8} \sin \left(\frac{4 \pi }{\omega_g}\right) & \frac{1}{4} \sin ^2\left(\frac{2 \pi }{\omega_g}\right) & \pi (1 -  g) \omega_g^{2/3}-\frac{1}{8} \sin \left(\frac{4 \pi }{\omega_g}\right) & \frac{1}{4} \sin ^2\left(\frac{2 \pi }{\omega_g}\right) \\
 \frac{1}{4} \sin ^2\left(\frac{2 \pi }{\omega_g}\right) & \frac{1}{8} \sin \left(\frac{4 \pi }{\omega_g}\right) & \frac{1}{4} \sin ^2\left(\frac{2 \pi }{\omega_g}\right) & \frac{\pi }{\omega_g}+\frac{1}{8} \sin \left(\frac{4 \pi }{\omega_g}\right) \\
\end{pmatrix} \nonumber
\end{equation}
which represent the covariance matrix of two initial squeezed thermal states entangling through a gaussian interaction $g (\hat{x}_1 - \hat{x}_2)^2$. While, for conciseness, we shall not give the specific form of the first moments, on the diagonal terms, they are found to be unchanged from the unitary case, while, in the off diagonal terms, the complex part is which leads to a different QRDM. The time evolution of the QRDM, omitting the time dependency of the contrasts and phases, is given by
\begin{equation}
   \hspace{-1.5cm} \hat{\varrho}_{q}  = \frac{1}{4}
\left(
\begin{array}{cccc}
 1 & 
 {\rm e}^{ -  \mathcal{C}_z -  \mathcal{C}_{s, n_p}^{(1)} - \mathcal{C}_{s,n_p}^{(2)} -  \mathcal{C}_{\Gamma}^{(1)} - \mathcal{C}_{\Gamma}^{(2)}  - i \phi_g  } & 
{\rm e}^{ -  \mathcal{C}_z - 
\mathcal{C}_{s,n_p}^{(1)} - \mathcal{C}_{s,n_p}^{(2)} -  \mathcal{C}_{\Gamma}^{(1)} - \mathcal{C}_{\Gamma}^{(2)}  - i \phi_g  } & 
 {\rm e}^{- 4 \mathcal{C}_z  -   4 \mathcal{C}_{s,n_p}^{(2)} -  4 \mathcal{C}_{\Gamma}^{(2)}   }
 \\
  {\rm e}^{-  \mathcal{C}_z  - \mathcal{C}_{s,n_p}^{(1)} - \mathcal{C}_{s,n_p}^{(2)} -  \mathcal{C}_{\Gamma}^{(1)} - \mathcal{C}_{\Gamma}^{(2)}  + i \phi_g  } & 
 1 &
 {\rm e}^{-  4 \mathcal{C}_z  -  4  \mathcal{C}_{s,n_p}^{(1)} -  4 \mathcal{C}_{\Gamma}^{(1)}   } & 
 {\rm e}^{ -  \mathcal{C}_z  - \mathcal{C}_{s,n_p}^{(1)} - \mathcal{C}_{s,n_p}^{(2)} -  \mathcal{C}_{\Gamma}^{(1)} - \mathcal{C}_{\Gamma}^{(2)}  + i \phi_g  } \\
  {\rm e}^{-  \mathcal{C}_z  - \mathcal{C}_{s,n_p}^{(1)} - \mathcal{C}_{s,n_p}^{(2)} -  \mathcal{C}_{\Gamma}^{(1)} - \mathcal{C}_{\Gamma}^{(2)}  + i \phi_g  } & 
 {\rm e}^{ -  4 \mathcal{C}_z - 4  \mathcal{C}_{s,n_p}^{(1)} -  4 \mathcal{C}_{\Gamma}^{(1)}   } & 
 1 & 
  {\rm e}^{ -  \mathcal{C}_z - \mathcal{C}_{s,n_p}^{(1)} - \mathcal{C}_{s,n_p}^{(2)} -  \mathcal{C}_{\Gamma}^{(1)} - \mathcal{C}_{\Gamma}^{(2)}  + i \phi_g  }  \\
 {\rm e}^{ - 4 \mathcal{C}_z - 4  \mathcal{C}_{s,n_p}^{(2)} - 4 \mathcal{C}_{\Gamma}^{(2)}   } & 
 {\rm e}^{-  \mathcal{C}_z  -  \mathcal{C}_{s, n_p}^{(1)} - \mathcal{C}_{s,n_p}^{(2)} -  \mathcal{C}_{\Gamma}^{(1)} - \mathcal{C}_{\Gamma}^{(2)}  - i \phi_g  } & 
  {\rm e}^{ -  \mathcal{C}_z  -  \mathcal{C}_{s, n_p}^{(1)} - \mathcal{C}_{s,n_p}^{(2)} -  \mathcal{C}_{\Gamma}^{(1)} - \mathcal{C}_{\Gamma}^{(2)}  - i \phi_g  } & 
 1 \\
\end{array}
\right) \nonumber
\end{equation}
where 
\begin{equation}
    \mathcal{C}_{s,n_p}^{(1)}(\tau) = (1+2 n_p) \frac{f_q^2}{4 \omega_g^4 s}  \left(\left(1-s^2 \omega_g^2\right) \cos (2 \tau  \omega_g)+s^2 \omega_g^2 -4 \cos (\tau  \omega_g)+3\right) \;,
\end{equation}
\begin{equation}
     \mathcal{C}_{s,n_p}^{(2)}(\tau) = (1+2 n_p) f_q^2 \sin ^2\left(\frac{\tau }{2}\right) \left(\left(s-\frac{1}{s}\right) \cos (\tau )+s+\frac{1}{s}\right) \;, 
\end{equation}
\begin{equation}
    \mathcal{C}_{\Gamma}^{(1)}(\tau) = \Gamma_x \frac{f_q^2 }{8 \omega_g^5} (6 \tau  \omega_g-8 \sin (\tau  \omega_g)+\sin (2 \tau  \omega_g))\; ,\hspace{0.5cm}
    \mathcal{C}_{\Gamma}^{(2)}(\tau) = \Gamma_x \frac{f_q^2 }{4}  (3 \tau +\sin (\tau ) (\cos (\tau )-4))\;,
\end{equation}
and $ \mathcal{C}_z = \Gamma_z \tau$, where the first two decays encode HD effects due to the superposition, the second two are the diffusion terms and the last one the dephasing term. We note that the diffusive decays (propositional to $\Gamma_x$) are independent on the number of phonons $n_p$ and squeezing parameter $s$. At $\tau_f$, it follows that $ \mathcal{C}_z = 2 \pi \Gamma_z$,
\begin{equation}
    \mathcal{C}_{s,n_p}^{(1)}(\tau_f) = 0 \hspace{0.5cm} \mathcal{C}_{s,n_p}^{(2)}(\tau_f) =  (1+2 n_p) f_q^2 \sin ^2\left(\frac{\pi }{\omega_g}\right) \left(\left(s-\frac{1}{s}\right) \cos \left(\frac{2 \pi }{\omega_g}\right)+s+\frac{1}{s}\right) 
\end{equation}
\begin{equation}
    \mathcal{C}_{\Gamma}^{(1)}(\tau_f) = \Gamma_x \frac{3 \pi f_q^2 }{2 \omega_g^5} \hspace{0.5cm}
    \mathcal{C}_{\Gamma}^{(2)}(\tau_f) = \Gamma_x \frac{f_q^2 }{4}  \left(\frac{6 \pi}{\omega_g} +\sin \left(\frac{2 \pi}{\omega_g}\right) \left(\cos \left(\frac{2 \pi}{\omega_g}\right)-4\right)\right)
\end{equation} 
such that the QRDM simplifies to 
\begin{equation}
   \hspace{-1cm} \hat{\varrho}_{q}(\tau_f)  = \frac{1}{4}
\left(
\begin{array}{cccc}
 1 & 
 {\rm e}^{ -\mathcal{C}_z  - \mathcal{C}_{s,n_p} -  \mathcal{C}_{\Gamma}^{(1)} - \mathcal{C}_{\Gamma}^{(2)}  - i \phi_g  } & 
{\rm e}^{-\mathcal{C}_z - \mathcal{C}_{s,n_p} -  \mathcal{C}_{\Gamma}^{(1)} - \mathcal{C}_{\Gamma}^{(2)}  - i \phi_g  } & 
 {\rm e}^{ -4 \mathcal{C}_z -  4 \mathcal{C}_{s,n_p} - 4  \mathcal{C}_{\Gamma}^{(2)}   }
 \\
  {\rm e}^{  -\mathcal{C}_z - \mathcal{C}_{s,n_p} -  \mathcal{C}_{\Gamma}^{(1)} - \mathcal{C}_{\Gamma}^{(2)}  + i \phi_g  } & 
 1 &
 {\rm e}^{ -\mathcal{C}_z - 4 \mathcal{C}_{\Gamma}^{(1)}   } & 
 {\rm e}^{ -\mathcal{C}_z  - \mathcal{C}_{s,n_p} -  \mathcal{C}_{\Gamma}^{(1)} - \mathcal{C}_{\Gamma}^{(2)}  + i \phi_g  } \\
  {\rm e}^{ -\mathcal{C}_z  - \mathcal{C}_{s,n_p} -  \mathcal{C}_{\Gamma}^{(1)} - \mathcal{C}_{\Gamma}^{(2)}  + i \phi_g  } & 
 {\rm e}^{ -\mathcal{C}_z -  4 \mathcal{C}_{s,n_p} - 4 \mathcal{C}_{\Gamma}^{(1)}   } & 
 1 & 
  {\rm e}^{  -\mathcal{C}_z - \mathcal{C}_{s,n_p} -  \mathcal{C}_{\Gamma}^{(1)} - \mathcal{C}_{\Gamma}^{(2)}  + i \phi_g  }  \\
 {\rm e}^{ -4 \mathcal{C}_z - 4  \mathcal{C}_{s,n_p} -  4 \mathcal{C}_{\Gamma}^{(2)}   } & 
 {\rm e}^{ -\mathcal{C}_z  - \mathcal{C}_{s,n_p} -  \mathcal{C}_{\Gamma}^{(1)} - \mathcal{C}_{\Gamma}^{(2)}  - i \phi_g  } & 
  {\rm e}^{ -\mathcal{C}_z  - \mathcal{C}_{s,n_p} -  \mathcal{C}_{\Gamma}^{(1)} - \mathcal{C}_{\Gamma}^{(2)}  - i \phi_g  } & 
 1 \\
\end{array}
\right) \nonumber
\end{equation}
Taking the trace with the witness operator given in Eq.~(\ref{eq:witness_approx}), the negativity reads
\begin{equation}
    \mathcal{N} =  \text{Tr}[\mathcal{W}\hat{\varrho}_{q}(\tau_f) ] = e^{ - \mathcal{C}_z  - \mathcal{C}_{s,n_p} -  \mathcal{C}_{\Gamma}^{(1)} - \mathcal{C}_{\Gamma}^{(2)}} \sin (\phi_g) - \frac{1}{4} \left(2 - e^{- 4  \mathcal{C}_{\Gamma}^{(1)} -4  \mathcal{C}_z } -  e^{- 4\mathcal{C}_z  - 4  \mathcal{C}_{s,n_p}  - 4 \mathcal{C}_{\Gamma}^{(2)}}
    \right)
\end{equation}

\section{Supplementary Plots \label{app:plots}}

\begin{figure}[H]
\includegraphics[width=1\textwidth]{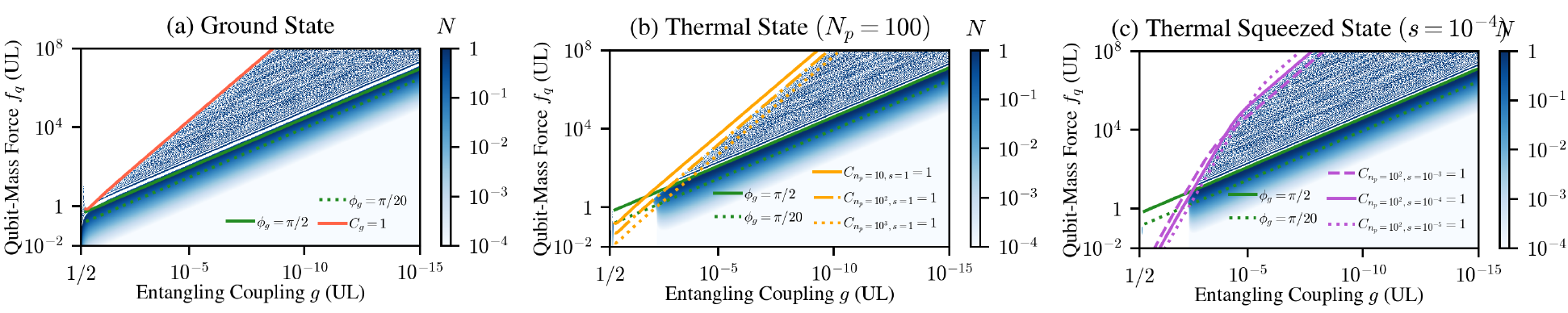}
  \caption{Negativity as function of the unitless parameter space for unitary dynamics with different initial states: \\
  (a) Ground state, (b) thermal states, and (c) squeezed thermal states ($n_p =100$).}% The common shift of the centers of the SG interferometers has been already considered.}
\end{figure}

\begin{figure}[H]
\includegraphics[width=1\textwidth]{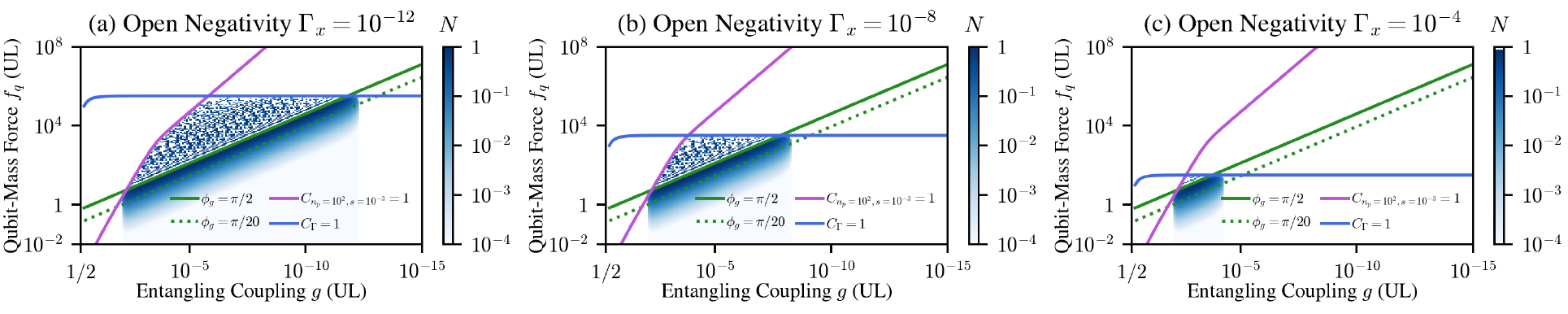}
  \caption{Negativity as function of the unitless parameter space for open dynamics with different diffision rates.}% The common shift of the centers of the SG interferometers has been already considered.}
\end{figure}

\twocolumngrid

\end{document}